\documentclass[aps,twocolumn,prd,superscriptaddress]{revtex4-2}

\usepackage[utf8]{inputenc}

\usepackage{mathtools}
\usepackage{amsfonts}
\usepackage{mathrsfs}
\usepackage{bbm}
\usepackage{slashed}

\usepackage{graphicx}
\usepackage{color}
\usepackage{array}

\usepackage{placeins}
\usepackage{booktabs}
\usepackage{makecell}
\usepackage{floatrow}
\usepackage{adjustbox}
\usepackage{tabularx}

\usepackage{xspace}
\usepackage{siunitx}
\usepackage{xfrac}
\usepackage{hyperref}
\usepackage[nameinlink]{cleveref}
\usepackage{appendix}
\usepackage{units}

\usepackage{xifthen}
\usepackage{xcolor}

\hypersetup{
	colorlinks,
	linkcolor={red!75!black},
	citecolor={blue!75!black},
	urlcolor={blue!75!black}
}

\setkeys{Gin}{width=0.48\textwidth}
\graphicspath{{./figures/}}

\newcommand{\gettitle}{QMeS-Derivation: \textit{Mathematica} package for the symbolic derivation of functional equations}

\hypersetup{
	pdftitle={\gettitle},
	pdfauthor={Pawlowski, Schneider, Wink},
	pdfkeywords={Yang-Mills theory} {Polyakov loop}
		{correlations functions} {temperature}
		{Debye mass} {confinement}  {deconfinement} 
		{functional renormalisation group}
		{background field} {order parameter},
	bookmarksopen=true,
	bookmarksopenlevel=2,
	bookmarksnumbered=true
}

\begin{document}

\title{\gettitle}

\author{Jan M. Pawlowski} 
\affiliation{Institut f\"ur Theoretische Physik,
	Universit\"at Heidelberg, Philosophenweg 16,
	69120 Heidelberg, Germany
}
\affiliation{ExtreMe Matter Institute EMMI,
	GSI, Planckstr. 1,
	64291 Darmstadt, Germany
}

\author{Coralie S. Schneider}
\affiliation{Institut f\"ur Theoretische Physik,
	Universit\"at Heidelberg, Philosophenweg 16,
	69120 Heidelberg, Germany
}

\author{Nicolas Wink}
\affiliation{Institut f\"ur Theoretische Physik,
	Universit\"at Heidelberg, Philosophenweg 16,
	69120 Heidelberg, Germany
}

\begin{abstract}
We present the \textit{Mathematica} package \textit{QMeS-Derivation}. It derives symbolic functional equations from a given master equation. The latter include functional renormalisation group equations, Dyson-Schwinger equations,  Slavnov-Taylor and Ward identities and their modifications in the presence of momentum cutoffs.  The modules allow to derive the functional equations, take functional derivatives, trace over field space, apply a given truncation scheme, and do  momentum routings while keeping track of prefactors and signs that arise from fermionic commutation relations. The package furthermore contains an installer as well as \textit{Mathematica} notebooks with showcase examples.
	\vspace*{55pt}
\end{abstract}

\maketitle

\section{Introduction}
\label{sec:Introduction}
Functional approaches are a well-established tool to study non-perturbative aspects of quantum field theories. They have been successfully used for a wide class of non-perturbative physics problems, ranging from strongly correlated condensed matter and statistical physics systems over nuclear physics, QCD and high energy physics to beyond the Standard Model physics, cosmology and quantum gravity. Applications also include real-time aspects in an out of equilibrium. For reviews on various physics applications of functional methods see e.g.\ \cite{Berges:2000ew, Polonyi:2001se,  Delamotte:2003dw, Pawlowski:2005xe, Schaefer:2006sr, Gies:2006wv, Delamotte:2007pf, Igarashi:2009tj, Rosten:2010vm, Kopietz:2010zz, Braun:2011pp, Litim:2011cp, Metzner_2012, Salmhofer:2018sgo, Eichhorn:2018yfc, Reuter:2019byg, Bonanno:2020bil,  Dupuis:2020fhh, Pawlowski:2020qer, Roberts:2000aa, Alkofer:2000wg, Fischer:2006ub, Binosi:2009qm, Maas:2011se,  Boucaud:2011ug, Aguilar:2015bud, Eichmann:2016yit, Sanchis-Alepuz:2017jjd, Huber:2018ned, Fischer:2018sdj}.

In these approaches one solves a set of functional  integro-differential loop relations between correlation functions of the theory at hand. These relations are typically closed at one or two-loop order in full correlation functions. If aiming for quantitative precision this requires setting up and solving a large set of loop equations involving the full tensor structure and momentum dependences of the correlation function involved. This requires the use of elaborate computer-algebraic tools as well as well-structured numerics. 

To date, there are still only a few computer-algebraic tools for functional methods~\cite{Alkofer:2008nt, Huber:2011qr, Huber_2020, github:DoFun, Vermaseren:2000nd, github:form, Feng:2012tk, github:FormLink, Cyrol:2016zqb, github:FormTracer, Litim:2020jvl, github:ARGES}. In this work we present a package of \textit{QMeS} (\textit{Quantum Master equations: environment for numerical Solutions}), that can be used for the symbolic derivation of functional equations arising from a master equation. Relevant examples are Functional Renormalisation Group (fRG) Equations, Dyson-Schwinger Equation (DSE) or Slavnov-Taylor Identities (STI) and their modification in the presence of a cutoff, the modified STIs (mSTI). In most cases the cutoff is an infrared cutoff, and hence the mSTI includes the STI as a special case for a vanishing cutoff. 

The package is written in \textit{Mathematica} and can be used to derive a functional equation such as fRGEs, DSEs, mSTIs from a given field content and, for the DSE, a given  classical action. Then, symbolic equations for different n-point functions,~i.e. the moments of the master equations, can be derived. Naturally, it works in a general field space, allowing for arbitrary theories and can include momentum routing for the diagrammatic/symbolic results. Its coherent implementation of conventions and handling of fermionic minus signs for diagrams allows for a simple and intuitive use. Due to its modular structure it facilitates future extensions to other master equations and more complicated objects and truncations. 

In the following, we introduce the master equation for fRG, DSE, and mSTI as well as our condensed notation. We proceed by describing the details of the package in \Cref{sec:description}, i.e. how the modules are connected via the interface, as well as the installation process. Then we give an overview of the input and output in \textit{QMeS-Derivation}. \Cref{sec:example} contains two examples: Yang-Mills and Yukawa theory ($N_f = 1$ and $N_f = 2$). For these example theories we describe, how to derive different symbolic functional equations from an action. In  \Cref{sec:Conclusion} we summarise the main features of \textit{QMeS-Derivation}.
\section{Quantum master equations}
\label{sec:conventions}
In this section we discuss the quantum master equation for the fRG, mSTI and DSE. Full derivations can be found in  \Cref{app:Derivation}. We use a superfield notation throughout the paper, introduced below.  

For a general quantum field theory the Euclidean action $S[\phi]$ reads  
\begin{align}
\label{eq:clAction}
	S[\phi] = \sum_{n \geq 2} \frac{1}{n!}S^{a_1 \dots a_n}\phi_{a_n} \dots \phi_{a_1}
\, ,
\end{align}
In \labelcref{eq:clAction} we have introduced deWitt's condensed notation, for the form used here see \cite{Pawlowski:2005xe}. The $a_i$ comprise internal and Lorentz indices, as well as species of fields and a sum/integration over space-time or momenta. The prefactor $1/n!$ is a vector-factorial, where each component corresponds to the factorial of the number of fields of the same species in the summand. Lowering and rising indices is done with the metric  $\gamma^{ab}$, that is diagonal in bosonic subspaces and symplectic in fermionic ones. For a single fermion anti-fermion pair $(f, \bar{f})$ the metric is given by,  
\begin{align}\label{eq:metric}
	(\gamma^{ab}) = \left( \begin{array}{cr}
		0 & -1 \\[1ex]
		1 & 0 
	\end{array}\right)\,. 
\end{align} 
For the complete metric we have the normalisation 
\begin{align}\nonumber 
	\gamma_b ^{\;a} &= \gamma^{ac}\gamma_{bc} = \delta^a_b\,,\nonumber\\[4pt]
	\gamma^a_{\;\,\,b} &= \gamma^{ac}\gamma_{cb} = (-1)^{ab}\delta^a_b\,,
\label{eq:inverse} \end{align}
with
\begin{align}
	(-1)^{ab} = \begin{cases}
		-1  &\text{$a$ and $b$ fermionic,}\\[4pt]
		\;\;\;1& \text{otherwise.}
	\end{cases}
	\label{eq:prefConvention}
\end{align}
Then, lowering and rising indices follows as, 
\begin{align}
	\phi_a = \phi^b \gamma_{ba}\,,\nonumber\\[4pt]
	\phi^a = \gamma^{ab}\phi_b\,.
\end{align}
The condensed notation introduced above allows us to write the Master equations in a concise form. 
Moreover, the metric introduced here is also used in the program.

The Schwinger functional $W[J]$, the generating functional of connected correlation functions with the classical action  \labelcref{eq:clAction}, follows as, 
\begin{align}
\label{eq:genFunc}
	e^{W[J]} = \int D \phi \exp\left(-S[\phi]+ J^a \phi_a\right) = Z[J]
\, .
\end{align}
In order to make the condensed notation more explicit, we write the source term as a sum over internal and 
Lorentz indices, and species of fields, $\alpha$, and a space-time integral,  
\begin{align}
	J^a\phi_a = \sum_\alpha \int d^d x\, J^\alpha(x)\phi_\alpha(x)
	\, .	
\end{align}
While the derivation of master equations is best done with the Schwinger functional and they also take the 
simplest form if formulated in $W[J]$, for a discussion see e.g.\ \cite{Pawlowski:2005xe}, most applications 
are done for the effective action $\Gamma[\Phi]$, the generating functional of one-particle irreducible (1PI) 
correlation functions. The argument of $\Gamma$ is the expectation value $\Phi$ of the field $\phi$,  
\begin{align}
	\frac{\delta W[J]}{\delta J^a} = W_a =  \left\langle \phi_a \right\rangle_J = \Phi_a
\, . \label{eq:Phi}
\end{align}
Then, the effective action $\Gamma[\Phi]$ is obtained as the Legendre transform of the Schwinger functional with respect to the source $J$, 
\begin{align}
\label{eq:effaction}
	\Gamma[\Phi] = \underset{J}{\sup}\Bigl(J^a\Phi_a-W[J]\Bigr)
\, .
\end{align}
\Cref{eq:effaction} entails that the sources are related to the derivatives of $\Gamma[\Phi]$ w.r.t.\ the fields, 
\begin{align}
	\frac{\delta \Gamma[\Phi]}{\delta \Phi_a}  = \Gamma^{a} = \gamma^a_{\;\,\,b}J^b
\, , 
\end{align}
where we have used 
\begin{align}
J^a\Phi_a = \Phi^a J_a = J_a \Phi^b \gamma^a_{\;\,\,b} = \Phi_b J^a \gamma^a_{\;\,\,b}\,.
\end{align}
Finally we are interested in master equations for correlation functions, provided by source- and field-derivatives of the Schwinger functional and the effective action respectively. We will use the notation, 
\begin{align}
\frac{\delta}{\delta \Phi_{a_1}} \dots \frac{\delta}{\delta \Phi_{a_n}}\Gamma[\Phi] &= \Gamma^{a_1\dots a_n}\,,\nonumber\\[2pt]
\frac{\delta}{\delta J^{a_1}} \dots \frac{\delta}{\delta J^{a_n}}W[J] &= W_{a_1\dots a_n}\,.
\label{eq:J-Phi-Derivatives}
\end{align}
The definition of the effective action entails, that the two-point function $\Gamma^{ab}$ is the inverse of the propagator $G_{ab}=W_{ab}$, 
\begin{align}
	\label{eq:Gab}
G_{ab}=\langle \phi_a \phi_b\rangle - \langle \phi_a \rangle\langle \phi_b\rangle\,.
\end{align}
In our condensed notation this reads, 
\begin{align}
G_{ac}\Gamma^{cb} = \gamma^b_{\;\,\,a}\,.
\label{propRelation}
\end{align}
With this setup we now derive quantum master equations in terms of the effective action, the flow equation for the effective action in the functional renormalisation group, the quantum equation of motion (Dyson-Schwinger equations), as well as the modified Slavnov-Taylor identities.

\subsection{Quantum equations of motion (DSE)}

\textit{Dyson-Schwinger equations} (DSEs)~\cite{Dyson:1949ha, Schwinger:1951ex} are the quantum equations of motion. They yield a complete description of the theory via 1PI correlation functions. In terms of the effective action they are given by, 
\begin{align}
\frac{\delta \Gamma[\Phi]}{\delta \Phi_a} = \frac{\delta S[\phi]}{\delta \phi_a}\Bigg|_{\phi_b = \Phi_b+G_{bc}\frac{\delta }{\delta \Phi_c}}\,.
\label{eq:DSE}
\end{align}
The full derivation of the DSE from the generating functional \labelcref{eq:genFunc} can be found in \Cref{app:DSEDerivation}. The r.h.s.\ 
of \labelcref{eq:DSE} comprises a classical part as well as loops. It is evident from a theory with an $n$th-order interaction of the fields 
leads to up to $n-2$-loops in full propagators, full vertices as well as one classical vertex. This entails, that the DSE is a closed (exact) $n-2$-loop functional master equation. As such it allows for perturbative as well as non-perturbative approximations. For reviews see e.g.\ \cite{Roberts:2000aa, Alkofer:2000wg, Fischer:2006ub, Binosi:2009qm, Maas:2011se,  Boucaud:2011ug, Aguilar:2015bud, Eichmann:2016yit, Sanchis-Alepuz:2017jjd, Huber:2018ned, Fischer:2018sdj}.

\subsection{Flow equation for the effective action (fRG)}
The flow equation for the effective action within the \textit{Functional Renormalisation Group} approach~\cite{Symanzik:1970rt,  Wetterich:1992yh, Ellwanger:1993mw, Bonini:1992vh, Morris:1993qb} can be viewed as a differential DSE. Typically, one introduces an (infrared) momentum regularisation, that suppresses quantum fluctuations below the infrared cutoff scale $k$. 
This is done by changing the classical dispersion by $1/2 \phi_a R^{ab}\phi_b$, where $R^{ab}$ is a momentum-dependent cutoff functions, that acts as a mass for low momenta and decays sufficiently fast for large momenta. More details and the derivation of the fRG equation, see \Cref{app:fRGderivation}. This approach allows us to integrate-out quantum fluctuations successively within momentum shells, finally arriving at the full effective action at $k=0$. With $\partial_t = k\partial_k$, the fRG flow equation is concisely given by, 
\begin{align}
\partial_t \Gamma =\frac{1}{2} \dot{R}^{ab} G_{ab}\, .
\label{eq:fRG}
\end{align}
\Cref{eq:fRG} is a one-loop exact master equation. The propagator $G_{ab}$ is infrared regularised via the cutoff mass. In turn, the equation is ultraviolet finite as $\dot{R}^{ab}$ decays for large momenta. In contradistinction to the DSE in \labelcref{eq:DSE}, that is $n-2$-loop exact for an $n$th order interaction, the fRG-master equation is a closed (exact) one-loop equation for general theories. As for the DSEs, a complete set of fRG-equations solves the theory exactly. For reviews see e.g.\ \cite{Berges:2000ew, Polonyi:2001se,  Delamotte:2003dw, Pawlowski:2005xe, Schaefer:2006sr, Gies:2006wv, Delamotte:2007pf, Igarashi:2009tj, Rosten:2010vm, Kopietz:2010zz, Braun:2011pp, Litim:2011cp, Metzner_2012, Salmhofer:2018sgo, Eichhorn:2018yfc, Reuter:2019byg, Bonanno:2020bil,  Dupuis:2020fhh, Pawlowski:2020qer}. 

\subsection{STI \& mSTI}\label{sec:STI+mSTI}

Within a gauge-fixed formulation of gauge theories the underlying gauge-invariance is carried by the BRST-symmetry (Becchi, Rouet, Stora, Tyutin). In their infinitesimal form, this symmetry is described by  the \textit{Slavnov-Taylor identities} (STI)~\cite{Becchi:1975nq,  Tyutin:1975qk}. They ensure the gauge invariance of observables, and can be formulated in terms of a master equation for the effective action including BRST-sources, \cite{ZinnJustin:1974mc, ZinnJustin:1999ze},   
\begin{align}
\frac{\delta \Gamma}{\delta Q^a} \frac{\delta \Gamma}{\delta \Phi_a} = 0\,. 
\label{eq:STI}
\end{align}
For deriving \Cref{eq:STI} one adds source terms $Q^a \mathfrak{s}\phi_a$ for BRST transformations to the path integral, for more details see \Cref{app:STIderivation}. The BRST transformation in Yang-Mills theory transforms a gauge boson into a ghost. The explicit transformation can be found in \Cref{eq:YMBRSTtrafo}. $Q^a$ is a source term for the BRST transformation of the fields $\mathfrak{s}\phi_a$, see \Cref{eq:BRSTgenFunc}. For reviews see e.g.\ \cite{Roberts:2000aa, Alkofer:2000wg, Pawlowski:2005xe, Fischer:2006ub, Binosi:2009qm, Maas:2011se,  Boucaud:2011ug, Aguilar:2015bud, Sanchis-Alepuz:2017jjd, Huber:2018ned}. 

However, the introduction of a cutoff term in the effective action breaks BRST symmetry for non-vanishing $k$. This leads to a modification of the symmetry identities (mSTI) that appears as a 1-loop correction, 
\begin{align}
\frac{\delta \Gamma}{\delta Q^a} \frac{\delta \Gamma}{\delta \Phi_a} = R^{ab}G_{bc}\Gamma^{c}_{\;\;Q^a}\,.
\label{eq:mSTI}
\end{align}
for more details see  \Cref{app:mSTIderivation}. In the above equation one can already see that for $k\rightarrow 0$ the mSTI reduces to the STI. Thus, satisfying the mSTI at all scales $k$, guarantees gauge invariance of observables at $k=0$. For details beyond that provided in \Cref{app:STIderivation} we refer to the reviews \cite{Pawlowski:2005xe, Gies:2006wv, Igarashi:2009tj, Rosten:2010vm, Reuter:2019byg, Bonanno:2020bil,  Dupuis:2020fhh, Pawlowski:2020qer} and references therein.


\section{Description}
\label{sec:description}

This section outlines the basic design and features of \textit{QMeS}, i.e. its modules and how they are connected via the interface. Furthermore we give instructions on how to install the package.
\subsection{Modules and Interface}

The code consists of four main modules - \textit{getDSE.m},  \textit{FunctionalDerivatives.m}, \textit{SuperindexDiagrams.m} and \textit{FullDiagrams.m} - which are connected by the interface \textit{DeriveFunctionalEquation.m}.

The four modules correspond to the four output options described in \Cref{sec:OutputOptions}. The workflow is depicted in \Cref{fig:workflowv0}.

The user is required to provide a setup that consists of either a master equation or an association indicating that \textit{QMeS} first needs to derive the DSE for a given classical action. Furthermore the setup needs to contain a definition of the field space and a truncation, as well as a list of field derivatives. Specifying the preferred form of the output (i.e. \texttt{"OutputLevel"}) is optional.

Depending on whether or not a master equation was provided the interface calls the \textit{FunctionalDerivatives.m} or first the \textit{getDSE.m} module which then generates the Dyson-Schwinger equation of the theory, and passes it on to the \textit{FunctionalDerivatives.m} module along with the setup and derivative list. Within this module the (remaining) field derivatives of the master equation are performed and fields are set to zero.

In the interface, the output and user provided input is again passed on to the \textit{SuperindexDiagrams.m} module, where the trace in field space is performed, the field content of objects, like propagators, n-point functions and regulator insertions, are sorted, prefactors are computed and the truncation is applied.

The result together with the initial input is then used by the \textit{FullDiagrams.m} module to replace the superfield indices with physical indices and the objects are replaced by functions of indices.

If the user has specified an output option, the workflow is terminated after the corresponding module providing the user with the chosen output. The default output option is \texttt{"FunctionalDerivatives"}.

\subsection{Requirements and Installation}

Functionality of \textit{QMeS-Derivation} is supported in \textit{Mathematica} 12.0 or higher, although it may also work with older versions.

To install the package download the installer via:
\begin{verbatim}
Import["https://raw.githubusercontent.com/
QMeS-toolbox/QMeS-Derivation/main/
QMeSInstaller.m"];
\end{verbatim}
Other options are to either save a copy of the repository in the \texttt{"../Mathematica/Applications"} folder or append the path (\texttt{yourpath}) where the copy is saved to the list of paths where \textit{Mathematica} searches for packages via:
\begin{verbatim}
AppendTo[$Path, "yourpath"];
\end{verbatim}
Then the package can be loaded in Windows by calling the following or an equivalent path for Linux and MacOS:
\begin{verbatim}
<<"QMeS-Derivation\\DeriveFunctionalEquation.m"
\end{verbatim}
\begin{figure*}[t]
	\includegraphics[width=1\textwidth]{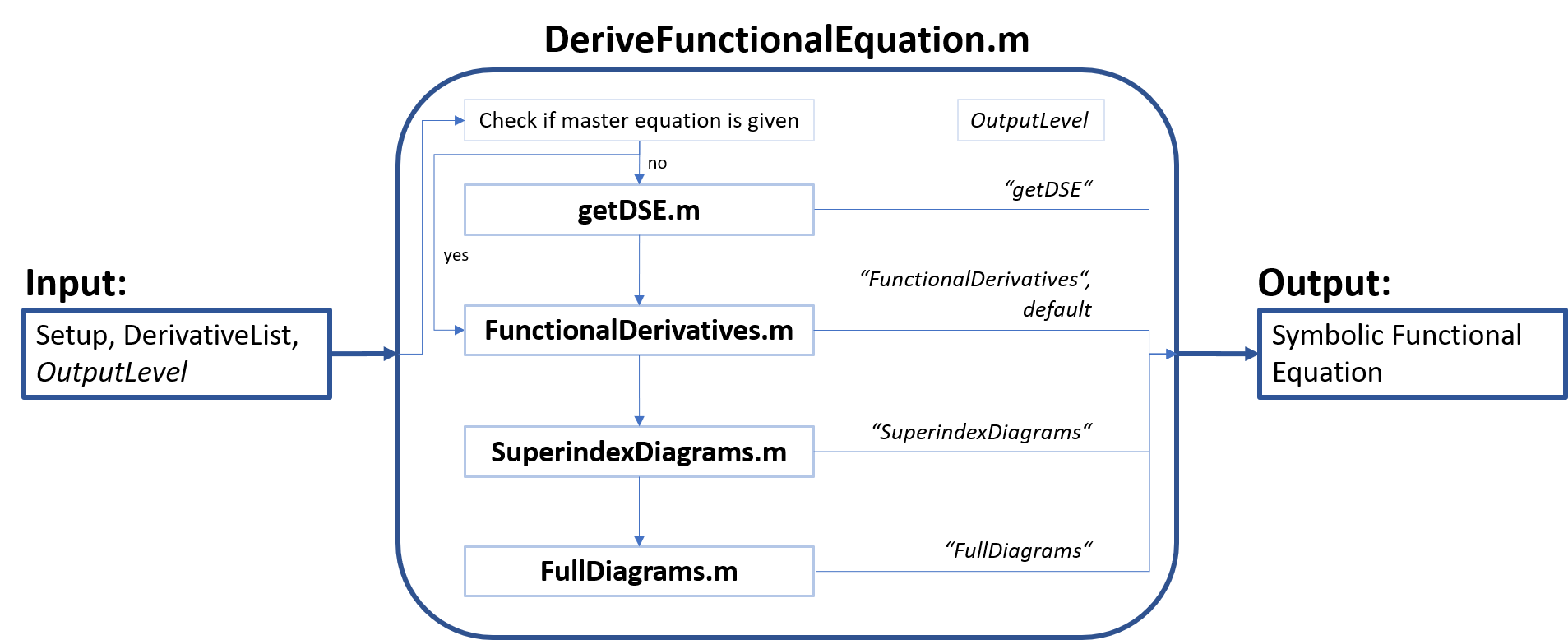}
	\caption{Depiction of the workflow of $\textit{QMeS-Derivation}$ with its interface and modules.}
	\label{fig:workflowv0}
\end{figure*}
%

\section{Input, Functions and Options}
\label{sec:functions}

To compute functional derivatives of a master equation one needs to define said equation as well as the theory one is working in. Both must be collected in an association.
\begin{verbatim}
Setup = <|"MasterEquation" -> masterEquation,
"FieldSpace" -> fields, 
"Truncation" -> truncation|>;
\end{verbatim}
If one first wants to derive a DSE of a given theory, the setup must be provided as,
\begin{verbatim}
SetupDSE = <|
"MasterEquation" -> <|"getDSE" -> "True", 
"classicalAction" -> classicalAction|>, 
"FieldSpace" -> fields, 
"Truncation" -> truncation|>;
\end{verbatim}
Note that one then needs a definition of the classical action via possible vertices.

\subsection{Master Equations and Objects}

Within the QMeS framework a master equation is defined as a list of objects, the first being an overall prefactor. Each object is of a specific \texttt{"type"} (e.g. propagator, n-point function, regulator or regulator derivative). Furthermore every object contains a list of \texttt{"indices"} that are superfield indices. For the fRG equation and the mSTI, the indices should be closed.
We recall the fRG \Cref{eq:fRG} as an example of a master equation,
\begin{align}
\partial_t \Gamma = \frac{1}{2} \dot{R}^{ab}G_{ab}\,, \nonumber
\end{align}
\begin{verbatim}
fRGEq = {"Prefactor" -> {1/2},
<|"type" -> "Regulatordot", 
"indices" -> {a, b}|>,
<|"type" -> "Propagator", 
"indices" -> {a, b}|>};
\end{verbatim}
as well as the modified Slavnov-Taylor identity (mSTI) introduced in \Cref{eq:mSTI},
\begin{align}
\frac{\delta \Gamma}{\delta Q^a}\frac{\delta \Gamma}{\delta \Phi_a} = R^{ab}G_{bc}\Gamma^c_{\;\;Q^a}\,. \nonumber
\end{align}
The mSTI can be written as:
\begin{verbatim}
LHSmSTIEq = {"Prefactor" -> {1}, 
<|"type" -> "nPoint", "indices" -> {Q[a]}, 
"nPoint" -> 1, "spec" -> "BRST"|>,
<|"type" -> "nPoint", "indices" -> {a}, 
"nPoint" -> 1, "spec" -> "none"|>};

mSTIEq = {"Prefactor" -> {1}, 
<|"type" -> "Regulator", "indices" -> {a, b}|>, 
<|"type" -> "Propagator", "indices" -> {b, c}|>, 
<|"type" -> "nPoint", "indices" -> {c, Q[a]},
	 "nPoint" -> 2, "spec" -> "BRST"|>};
\end{verbatim}
It is furthermore possible to derive the DSE of a given theory with the aforementioned setup. For further information see section \Cref{sec:OutputOptions}. The superindices in the master equations should not coincide with names of fields or any of their indices.

\subsubsection*{\textbf{\textup{Prefactors}}}

The first entry in every diagram is the \texttt{Prefactor}. It can contain numbers ($1$,$-1$,$1/2$,...) or a metric factor $(-1)^{ab}$. For example the prefactor
\begin{verbatim}
"Prefactor" -> {-1/2, {a,b}, {b,b}, {b,c}};
\end{verbatim} 
translates into
\begin{align}
-\frac{1}{2}(-1)^{ab}(-1)^{bb}(-1)^{bc}\,,
\end{align}
where again the superfield index convention introduced in \Cref{eq:prefConvention} is used.

\subsubsection*{\textbf{\textup{Regulator and Regulator Derivative}}}

\begin{verbatim}
<|"type" -> "Regulatordot", 
"indices" -> {a, b}|>;

<|"type" -> "Regulator", "indices" -> {a, b}|>;
\end{verbatim}
A regulator $R^{ab}$ or regulator derivative $\dot{R}^{ab}$ is an object with two superfield indices corresponding to the incoming and outgoing fields with their respective momenta and indices.

\subsubsection*{\textbf{\textup{Propagator}}}

\begin{verbatim}
<|"type" -> "Propagator", "indices" -> {a, b}|>;
\end{verbatim}
A propagator $G_{ab}$ is an object with two superfield indices corresponding to the fields and their indices. These are lower indices. Note that for fRG and mSTI equations the propagator is $k$-dependent whereas it is not for DSEs.

\subsubsection*{\textbf{\textup{n-Point Functions}}}

\begin{verbatim}
<|"type" -> "nPoint", "indices" -> {a, b, c, d},
"nPoint" -> 4, "spec" -> "none"|>;

<|"type" -> "nPoint", "indices" -> {a, b},
"nPoint" -> 2, "spec" -> "classical"|>;

<|"type" -> "nPoint", "indices" -> {a, b, Q[c]},
"nPoint" -> 3, "spec" -> "BRST"|>;
\end{verbatim}
n-Point functions are field derivatives of the effective action. The value of \texttt{"nPoint"} indicates the number of derivatives, whereas the \texttt{"indices"} again represent the superfield indices. The specification \texttt{"spec"} implies whether the vertex is a BRST (\texttt{"BRST"}, $\Gamma^{ab}_{\;\;\;Q^c}$), a 1PI (\texttt{"none"}, $\Gamma^{abcd}$) or a classical (\texttt{"classical"}, $S^{ab}$) one. The superfield index of a BRST source needs to be written as \texttt{"Q[i]"} to indicate that this is a lower index belonging to the BRST source of a field \texttt{Q[field]} (for the notation, see \Cref{sec:fields}). Again it is worth mentioning, that in case of fRG or mSTI equations the 1PI and BRST vertices are $k$-dependent objects.

\subsubsection*{\textbf{\textup{Fields}}}

\begin{verbatim}
<|"type" -> "Field", "indices" -> {a}|>;
\end{verbatim}
Fields $\Phi_a$ are objects with one lower index. Note that after taking all functional derivatives, external fields, which are left over, are set to zero.

\subsection{Theory}

The user is required to define a specific theory. This breaks down into two main parts: defining the fields with the respective indices and the truncation.

\subsubsection{Fields with indices}
\label{sec:fields}
The fields of a theory are either fermionic or bosonic. Antifermion/fermion pairs must be combined in a list. 
\begin{verbatim}
fields = 
<|"bosonic" -> {A[p, {mu, a}], B[p]},

"fermionic" -> {{cbar[p, {a}], c[p, {a}]},
				{af[p,{d}], f[p,{d}]}},
				
"BRSTsources" -> 
{{Q[A], "fermionic"}, {Q[B], "fermionic"}, 
{Q[cbar], "bosonic"}, {Q[c], "bosonic"},
{Q[af], "bosonic"}, {Q[f], "bosonic"}}|>;
\end{verbatim}
If a theory contains no fields of either bosonic or fermionic statistics, it is then required to assign an empty list.

When computing mSTIs one also needs to define the BRST charges of fields. They are indicated by \texttt{Q[field]} followed by the respective property of the charge (either \texttt{"fermionic"} or \texttt{"bosonic"}). For the computation of DSE or fRG equations it is not necessary to define the BRST sources.

The respective indices are provided as arguments of the fields, where the momentum is always the first entry, followed by a list of further indices (e.g. group or Lorentz indices). Note that the names of the indices for different fields does not need to be unique. For better readability it is recommended to define the same kind of index with the same name: these names (e.g.\ \texttt{\{mu, i\}}) in combination with a unique number (e.g.\ \texttt{\$8215}) will be used to create unique indices (e.g.\ \texttt{\{mu\$8215, i\$8215\}}) by \textit{QMeS}.

\subsubsection{Truncation and classical Action}

For the derivation of DSEs it is necessary to define the classical action via vertices. This is done by giving a list of combination of fields that appear as a classical vertex in the action,
\begin{verbatim}
classicalAction = {{A, A}, {c, cbar}, {A, A, A},
{A, A, A, A}, {A, c, cbar}};
\end{verbatim}
Furthermore the truncation of the full theory is defined by specifying the truncation of 1PI and BRST vertices. It is worth mentioning that the user is also required to include the possible propagators in this list,
\begin{verbatim}
Truncation = {{A, A}, {c, cbar}, {A, A, A}, 
{A, A, A, A}, {A, c, cbar}, {A, A, c, cbar}};
\end{verbatim}
The truncation may be similar or include more vertices than the classical action.

In both definitions the order of fields or vertices is irrelevant.

\subsection{Derivative List}

Lastly one needs to specify a list of field derivatives. Note that the last entry of the list will be the first derivative. 
\begin{verbatim}
DerivativeList1 = {A, A};
DerivativeList2 = {A[a], A[b]};
DerivativeList3 = 
               {A[-p, {mu, a}], A[p, {nu, b}]};
\end{verbatim}
Generally one has three options: the first is to only provide the field names. This can be combined with the output options \texttt{"getDSE"} and \texttt{"FunctionalDerivatives"}. The second is to assign superindices to the fields, this input additionally works with \texttt{"SuperindexDiagrams"}. If one wants to obtain full diagrams with momentum routing (\texttt{"FullDiagrams"}), then one needs to assign indices and momenta to the fields.

\subsection{Outputs}
\label{sec:OutputOptions}

The main function takes two arguments: the setup and the list of field derivatives,
\begin{verbatim}
DeriveFunctionalEquation[Setup, DerivativeList];
\end{verbatim}
The output is always a list of the diagrams that are produced. The specification of the diagrams can be altered with \texttt{options}.

Options are called via
\begin{verbatim}
DeriveFunctionalEquation[Setup, DerivativeList, 
"OutputLevel" ->  options];
\end{verbatim}
There are three options specifying the level of the output via \texttt{"OutputLevel"}:

\subsubsection*{\textbf{\textup{getDSE}}}

The first option is to simply derive the Dyson-Schwinger equation via \texttt{getDSE}. The user needs to specify the classical vertices as well as at least one field derivative $\frac{\delta}{\delta \phi}$. From this the RHS of the DSE is computed according to the rules in \Cref{app:DerRules}. This means the classical action \labelcref{eq:clAction} can be written as,
\begin{align}
S[\phi] &= \frac{1}{2!}S^{a_1a_2}\phi_{a_2}\phi_{a_1} + \frac{1}{3!}S^{a_1a_2a_3}\phi_{a_3}\phi_{a_2}\phi_{a_1}\nonumber\\
&+\dots+\frac{1}{n!}S^{a_1\dots a_n}\phi_{a_n}\dots\phi_{a_1}\,,
\end{align}
where only those orders appear that are given in the theory and where we have restricted ourselves for the sake of simplicity to only one species of fields.  
The \textit{getDSE} module then computes
\begin{align}
\frac{\delta S}{\delta \phi_i}\bigg\rvert_{\phi_i = \Phi_i +G_{ij}\frac{\delta}{\delta \Phi_j}}\,.
\end{align}
Terms that end with a field derivative are immediately dropped. One then obtains the general diagrams that contribute to the DSE for a given classical action. If the field derivative list contains more than one entry, the last one is a field, whereas the others are processed as expectation values of fields.
\begin{verbatim}
DerivativeListDSE = 
{Phi[a], Phi[b], Phi[c], Phi[d]}
\end{verbatim}
The list above thus produces
\begin{align}
\Gamma^{abcd} = \frac{\delta^3}{\delta \Phi_a\Phi_b\Phi_c} \left(\frac{\delta S}{\delta \phi_d}\right)_{\phi_i = \Phi_i +G_{ij}\frac{\delta}{\delta \Phi_j}}\,.
\end{align}

\subsubsection*{\textbf{\textup{FunctionalDerivatives}}}

If the option is set to \texttt{"FunctionalDerivatives"} the user obtains a list of diagrams that are generated by taking functional derivatives of the quantum master equation. The trace over fields in the diagrams is however not taken. Therefore one gets symbolic diagrams with fields set to zero. When choosing this option with a given master equation, the user is not required to specify a truncation.

The default output level is equal to calling the \texttt{"FunctionalDerivatives"} option. 

\subsubsection*{\textbf{\textup{SuperindexDiagrams}}}

The third option is called \texttt{"SuperindexDiagrams"}. If this is chosen, the trace over fields is taken, where only those diagrams remain that satisfy the truncation, and the fields in the objects are sorted canonically. This means that upper indices (eg. for regulators, regulator derivatives or vertices, including BRST-vertices) are sorted as (\textit{bosonic}, \textit{antifermionic}, \textit{fermionic}) and lower fermionic indices in reverse order. If two indices have are of the same type, e.g.\ two bosonic fields, they are sorted alphabetically. Lastly the prefactors are evaluated. 

\textit{QMeS} aims at generating outputs for general theories. For this reason we refrain from symmetrization or identification procedures for diagrams.

\subsubsection*{\textbf{\textup{FullDiagrams}}}

For the last module one may call the main function with the option \texttt{"FullDiagrams"}, which means that in addition to the previous steps also the momentum routing is done for all 1-loop diagrams (i.e. fRG, mSTI, but not for all DSE diagrams). Superfield indices are replaced by physical indices and objects are transformed into functions of indices such that one can insert Feynman rules easily.

\section{Examples}
\label{sec:example}

In this section we give different examples of deriving symbolic functional equations with \textit{QMeS}.

The first example is deriving functional equations (i.e. fRG, mSTI and DSE) within Yang-Mills theory which serves as a prerequisite for QCD. Studying QCD with functional methods is an ab initio approach to investigate the non-perturbative regime.

Then we derive fRG equations in $N_f=1$ and $N_f=2$ Yukawa theory. It illustrates and emphasizes how \textit{QMeS} handles multiple fermions and sorts the vertices accordingly. Furthermore a simple Yukawa model can already be used to describe nuclear forces between fermions which are mediated by pions thus approximating QCD with an effective field theory.

\subsection{Yang-Mills theory}
\label{sec:YangMills}

In the following we want to give the crucial steps one needs to take to compute functional equations in Yang-Mills theory (YM) with \textit{QMeS}.

The theory we work in is SU(3) Yang-Mills theory, thus one has bosonic gauge fields $A_{\mu}^a(p)$, fermionic ghosts $c^a(p)$ and antighosts $\bar{c}^a(p)$.
The classical Euclidean YM action including gauge fixing and ghost terms can be written as,
\begin{align}
S = \int d^4x &\left(\frac{1}{4} F^a_{\mu\nu}F^a_{\mu\nu}+\frac{1}{2\alpha}\partial_\mu A^a_\mu\partial_\nu A^a_\nu\right.\nonumber\\
&\left.+\partial_\mu \bar{c}^a \left(\partial_\mu c^a+g f^{abc} A^b_\mu c^c\right)\right)\,,
\end{align} 
with $F^a_{\mu\nu} = \partial_\mu A^a_\nu-\partial_\nu A^a_\mu +g f^{abc}A^b_\mu A^c_\nu\,$.
After Legendre transforming the classical action and introducing a regulator term one obtains the effective average action,
\begin{align}
\Gamma = &\frac{1}{2}\Gamma^{AA}AA-\Gamma^{\bar{c}c}\bar{c}c-\Gamma^{A\bar{c}c}A\bar{c}c\nonumber\\[4pt]
&+\frac{1}{6}\Gamma^{AAA}AAA+\frac{1}{24}\Gamma^{AAAA}AAAA\nonumber\\[4pt]
&-\frac{1}{2}\Gamma^{AA\bar{c}c}AA\bar{c}c+\frac{1}{4}\Gamma^{\bar{c}\bar{c}cc}\bar{c}\bar{c}cc\,,
\end{align}
with indices suppressed. For a BRST-symmetric action one includes the source terms,
\begin{align}
\Gamma_{BRST} = &-\Gamma^{c}_{\;Q^A}cQ^A+\Gamma^{A}_{\;Q^{\bar{c}}}AQ^{\bar{c}}\nonumber\\[4pt]
&-\Gamma^{Ac}_{\;\;\;Q^A}AcQ^A-\frac{1}{2}\Gamma_{Q^c}^{\;\;\;\;cc}Q^ccc\,,
\end{align}
which relate to the BRST transformations,
\begin{align}
\mathfrak{s}A_\mu^a &= \partial_\mu c^a+g f^{abc} A_\mu^b c^c \delta \lambda\nonumber\\[4pt]
\mathfrak{s}c^a &= \frac{1}{2} g f^{abc}c^bc^c  \delta \lambda\nonumber\\[4pt]
\mathfrak{s}\bar{c}^a &= -\frac{1}{\alpha} \partial_\mu A_\mu^a  \delta \lambda\,,
\label{eq:YMBRSTtrafo}
\end{align}
with the infinitesimal transformation parameter $ \delta \lambda$ via $\left< \mathfrak{s} \phi_a\right> = -\frac{\delta \Gamma}{\delta Q^a}$. For more details see \Cref{app:STIderivation} and \Cref{app:mSTIderivation}. 

For pure Yang-Mills theory we can define the fields in \textit{QMeS} as,
\begin{verbatim}
fieldsYM = 
<|"bosonic" -> {A[p, {mu, a}]},
"fermionic" -> {{cbar[p, {a}], c[p, {a}]}}|>;
\end{verbatim}
Note that fermions need to be defined as a pair of the antifermion and corresponding fermion.

\begin{figure*}[t]
	\includegraphics[width=0.9\textwidth]{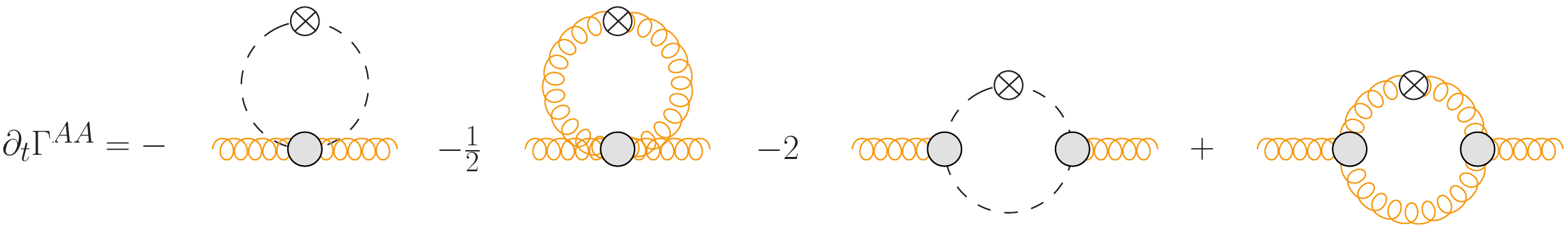}
	\caption{Graphical representation of the flow equation of the gluon two-point function, explicitly given in \labelcref{eq:flow_gluon_twopoint}. The dashed lines represent the ghost, curly orange lines the gluon. Full, grey blobs represent full vertices and the crossed circle represents the regulator derivative.}
	\label{fig:frg_gluon_fig}
\end{figure*}

Next we specify the truncation for the effective average action without the BRST terms. It is important to also define the two-point functions in order to get the possible propagators.
\begin{verbatim}
TruncationYM = {{A, A}, {c, cbar}, {A, A, A}, 
{A, A, A, A}, {A, c, cbar}, {A, A, c, cbar},
{c, c, cbar, cbar}};
\end{verbatim}
The classical Yang-Mills action is given by
\begin{verbatim}
classicalActionYM = {{A, A}, {c, cbar}, 
{A, A, A}, {A, A, A, A}, {A, c, cbar}};
\end{verbatim}
Since we have a theory with ghosts $c^a$ and color indices $a,\, b,\, d,\,\dots$, we use $i,\, j,\, m\,\dots$ as superindices for the master equations.
\subsubsection{Flow of the gluon two-point function}
To compute the flow of the gluon two-point function we need to define the Quantum Master equation which is in this case the fRG equation \labelcref{eq:fRG}. This translates to \textit{QMeS} input as,
\begin{verbatim}
fRGEq = {"Prefactor" -> {1/2},
<|"type" -> "Regulatordot", 
"indices" -> {i, j}|>,
<|"type" -> "Propagator", 
"indices" -> {i, j}|>};
\end{verbatim}
Now we can define the setup 
\begin{verbatim}
SetupYMfRG = <|"MasterEquation" -> fRGEq,
"FieldSpace" -> fieldsYM, 
"Truncation" -> TruncationYM|>;
\end{verbatim}
The only thing that is missing is a specification of the field derivatives that we want to take:
\begin{verbatim}
DerivativeListAA = 
{A[-p, {mu, a}], A[p, {nu, b}]};
\end{verbatim}
Now we can derive symbolic diagrams. In general we have different output options (see \Cref{sec:OutputOptions}).

First we can take a look at the general structure of diagrams that are produced when taking two functional derivatives with respect to the superfields $\Phi_{a}$ and $\Phi_{b}$ by calling the \textit{QMeS} command \texttt{DeriveFunctionalEquation} with the output option \texttt{"OutputLevel" -> "FunctionalDerivatives"}. We then obtain
\begin{align}
\label{eq:flow_gluon}
\dot{\Gamma}^{ab} =&-\frac{1}{2}(-1)^{ia}(-1)^{ib}(-1)^{nn}\dot{R}^{ij}G_{im}\Gamma^{mabn}G_{nj}\nonumber\\[4pt]
&+\frac{1}{2}(-1)^{ia}(-1)^{ib}(-1)^{nn}(-1)^{n'n'}\nonumber\\
&\dot{R}^{ij}G_{im}\Gamma^{man}G_{nm'}\Gamma^{m'bn'}G_{n'j}\nonumber\\[4pt]
&+\frac{1}{2}(-1)^{ia}(-1)^{ib}(-1)^{nn}(-1)^{n'n'}(-1)^{ab}\nonumber\\
&\dot{R}^{ij}G_{im}\Gamma^{mbn}G_{nm'}\Gamma^{m'an'}G_{n'j}\,.
\end{align}
One thus gets a tadpole diagram and two diagrams with two three-point vertices respectively.
Next we want to get the fully traced diagrams by evaluating
\begin{verbatim}
fRGDiagramsAA = DeriveFunctionalEquation[
SetupYMfRG, DerivativeListAA, 
"OutputLevel" -> "FullDiagrams"];
\end{verbatim} 
\begin{figure*}[t]
	\includegraphics[width=0.9\textwidth]{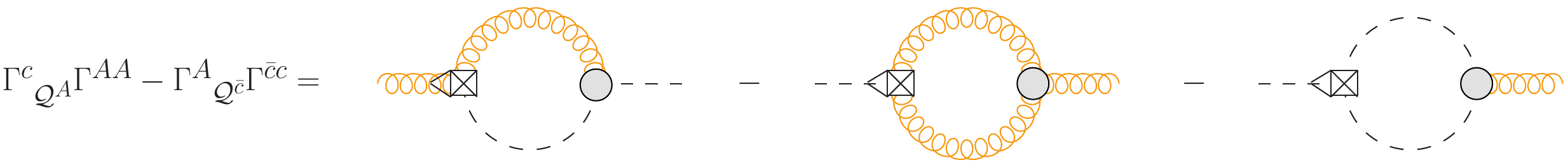}
	\caption{Graphical representation of the mSTI of the gluon two-point function, explicitly given in \labelcref{eq:mSTITwoPoint}. The dashed lines represent the ghost, curly orange lines the gluon. Full, grey blobs represent full vertices and the crossed square with the triangle the contraction of a regulator with a BRST vertex.}
	\label{fig:msti_gluon}
\end{figure*}

As a result we obtain in superindex notation where now $a \simeq (-p, \mu, a)$ and $b \simeq (p, \nu, b)$,
\begin{align}
\label{eq:flow_gluon_twopoint}
\dot{\Gamma}^{A_aA_b} =&-\dot{R}^{\bar{c}c}G_{c\bar{c}}\Gamma^{A_aA_b\bar{c}c}G_{c\bar{c}}\nonumber\\[4pt]
&-\frac{1}{2}\dot{R}^{AA}G_{AA}\Gamma^{AA_aA_bA}G_{AA}\nonumber\\[4pt]
&+\frac{1}{2}\dot{R}^{AA}G_{AA}\Gamma^{AA_aA}G_{AA}\Gamma^{AA_bA}G_{AA}\nonumber\\[4pt]
&+\frac{1}{2}\dot{R}^{AA}G_{AA}\Gamma^{AA_bA}G_{AA}\Gamma^{AA_aA}G_{AA}\nonumber\\[4pt] 
&-\dot{R}^{\bar{c}c}G_{c\bar{c}}\Gamma^{A_a\bar{c}c}G_{c\bar{c}}\Gamma^{A_b\bar{c}c}G_{c\bar{c}}\nonumber\\[4pt]
&-\dot{R}^{\bar{c}c}G_{c\bar{c}}\Gamma^{A_b\bar{c}c}G_{c\bar{c}}\Gamma^{A_a\bar{c}c}G_{c\bar{c}}\,.
\end{align} 
The \textit{QMeS} output is a list of different traced diagrams such that one can easily define and insert the Feynman rules for the different objects like propagators, regulators or vertices. It can be found in \Cref{app:fRGappendix}. A graphical representation of the flow can be found in \Cref{fig:frg_gluon_fig}.

\subsubsection{mSTI of gluon two-point function}
To compute the mSTI of the gluon two-point function we need to alter our definition of fields and include the corresponding BRST sources.
\begin{verbatim}
fieldsYMmSTI = <|"bosonic" -> {A[p, {mu, a}]},
"fermionic" -> {{cbar[p, {a}], c[p, {a}]}},
"BRSTsources" -> {{Q[A], "fermionic"}, 
{Q[cbar], "bosonic"}, {Q[c], "bosonic"}}|>;
\end{verbatim}

The truncation then also changes. The vertices on the right-hand side of the mSTI are for the sake of simplicity truncated as,
\begin{verbatim}
TruncationYMRHSmSTI = {{A, A}, {c, cbar},
{A, A, A}, {A, A, A, A}, {A, c, cbar}, 
{A, c, Q[A]}, {c, c, Q[c]}};
\end{verbatim}
and for the left-hand side we choose
\begin{verbatim}
TruncationYMLHSmSTI = {{A, A}, {c, cbar}, 
{A, A, A}, {A, A, A, A}, {A, c, cbar}, 
{A, Q[cbar]}, {c, Q[A]}, {A, c, Q[A]}, 
{c, c, Q[c]}};
\end{verbatim}
Lastly we need to define the right- and left-hand side of the mSTI equation \labelcref{eq:mSTI}. In the \textit{QMeS} formalism this is done by
\begin{verbatim}
mSTIRHS = {"Prefactor" -> {1}, 
<|"type" -> "Regulator", "indices" -> {i, j}|>, 
<|"type" -> "Propagator", "indices" -> {j, m}|>, 
<|"type" -> "nPoint", "indices" -> {m, Q[i]}, 
"nPoint" -> 2, "spec" -> "BRST"|>};

mSTILHS = {"Prefactor" -> {1}, 
<|"type" -> "nPoint", "indices" -> {Q[i]}, 
"nPoint" -> 1, "spec" -> "BRST"|>,
<|"type" -> "nPoint", "indices" -> {i}, 
"nPoint" -> 1, "spec" -> "none"|>};
\end{verbatim}
We define the two setups as,
\begin{verbatim}
SetupYMmSTIRHS = <|"MasterEquation" -> mSTIRHS,
"FieldSpace" -> fieldsYMmSTI, 
"Truncation" -> TruncationYMRHSmSTI|>;

SetupYMmSTILHS = <|"MasterEquation" -> mSTILHS,
"FieldSpace" -> fieldsYMmSTI, 
"Truncation" -> TruncationYMmSTILHS|>;
\end{verbatim}
To obtain the mSTI of the gluon two-point function one needs to take derivatives with respect to the ghost and gluon field,
\begin{verbatim}
DerivativeListAAmSTI = 
{A[-p, {mu, a}], c[p, {b}]};
\end{verbatim}
We obtain the full mSTI by evaluating
\begin{verbatim}
mSTIDiagramsAALHS = DeriveFunctionalEquation[
SetupLHSmSTILHS, DerivativeListmSTI, 
"OutputLevel" -> "FullDiagrams"];

mSTIDiagramsAARHS = DeriveFunctionalEquation[
SetupmSTIRHS, DerivativeListmSTI, 
"OutputLevel" -> "FullDiagrams"];
\end{verbatim}
With the superindices $a \simeq (-p,\mu,a)$ and $b\simeq (p,b)$ the algebraic equations are then given as,
\begin{align}
\Gamma^{c_b}_{\;\;Q^A}\Gamma^{AA_a}&-\Gamma^{A_a}_{\;\;\;\;Q^{\bar{c}}}\Gamma^{\bar{c}c_b} = R^{AA}G_{AA}\Gamma^{A\bar{c}c_b}G_{c\bar{c}}\Gamma^{A_ac}_{\;\;\;\;\;\;Q^A}\nonumber\\[4pt]
&-R^{AA}G_{AA}\Gamma^{AA_aA}G_{AA}\Gamma^{Ac_b}_{\;\;\;\;\;Q^A}\nonumber\\[4pt]
&-R^{\bar{c}c}G_{c\bar{c}}\Gamma^{A_a\bar{c}c}G_{c\bar{c}}\Gamma_{Q^c}^{\;\;\;\;cc_b}\,,
\label{eq:mSTITwoPoint}
\end{align}
where for the sake of brevity, indices and momenta are dropped. The output of \textit{QMeS} is given in \Cref{app:mSTIappendix}. The diagrams that contribute to the mSTI can be found in \Cref{fig:msti_gluon}.

\begin{figure*}[t]
	\includegraphics[width=0.9\textwidth]{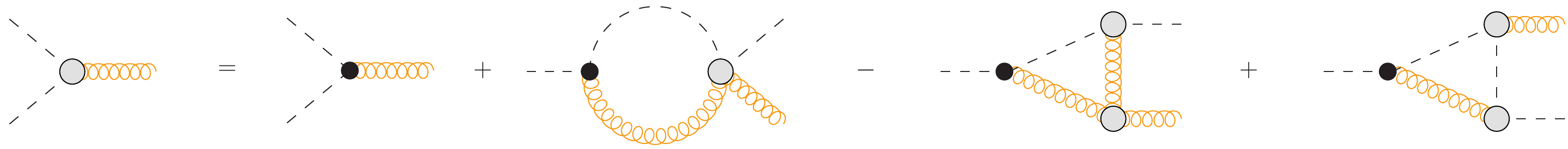}
	\caption{Graphical representation of the DSE of the ghost-gluon vertex, explicitly given in \labelcref{eq:dse_ghost_ghost_gluon}. The dashed lines represent the ghost, curly orange lines the gluon. Full, grey blobs represent full vertices and small, black blobs represent classical vertices.}
	\label{fig:dse_ghost_ghost_gluon}
\end{figure*}
%

\subsubsection{DSE of ghost-gluon vertex}

In this subsection we derive the DSE for the ghost-gluon vertex. This can be done by taking functional derivatives of the action,
\begin{align}
\Gamma^{A\bar{c}c} = \frac{\delta^2}{\delta A \delta \bar{c}} \left(\frac{\delta S}{\delta c}\right)_{\phi_a\rightarrow \Phi_a+G_{ab}\frac{\delta}{\delta \Phi_b}}\,.
\end{align}
We define the setup,
\begin{verbatim}
SetupYMDSE = <|"MasterEquation" -> 

<|"getDSE" -> "True", 
"classicalAction" -> classicalActionYM|>,

"FieldSpace" -> fieldsYM, 
"Truncation" -> TruncationYM|>;
\end{verbatim}
We define the derivative list
\begin{verbatim}
DerivativeListAcbarcDSE = {A[p1, {mu, a}], 
cbar[p2, {b}], c[-p1 - p2, {d}]};
\end{verbatim}
We get the full result by using the command:
\begin{verbatim}
DSEDiagramsAcbarc = DeriveFunctionalEquation[
SetupYMDSE, DerivativeListAcbarcDSE, 
"OutputLevel" -> "FullDiagrams"];
\end{verbatim} 
Diagrammatically the result is, 
\begin{align}
\label{eq:dse_ghost_ghost_gluon}
\Gamma^{A_a\bar{c}_bc_d} = &\;\;\;\;S^{A_a\bar{c}_bc_d}\nonumber\\[4pt]
&+S^{A\bar{c}c_d}G_{AA}\Gamma^{AA_a\bar{c}_bc}G_{c\bar{c}}\nonumber\\[4pt] &-S^{A\bar{c}c_d}G_{AA}\Gamma^{AAA_a}G_{AA}\Gamma^{A\bar{c}_bc}G_{c\bar{c}}\nonumber\\[4pt]
&+S^{A\bar{c}c_d}G_{AA}\Gamma^{A\bar{c}_bc}G_{c\bar{c}}\Gamma^{A_a\bar{c}c}G_{c\bar{c}}\,,
\end{align}
where we have used the superindices $a \simeq (p1,\mu,a)$, $b\simeq (p2,b)$ and $d \simeq (-p1-p2,d)$. The full equation can be found in \Cref{app:DSEAppendix}. The symbolic DSE can be found in \Cref{fig:dse_ghost_ghost_gluon}.

\subsection{Yukawa theory}

In this example we want to compute simple two-point flows in Yukawa theory. To further illustrate how \textit{QMeS} handles multiple fermions we do this in $N_f=1$ as well as $N_f=2$. Generally we can write the action of a Yukawa theory as,
\begin{align}
S &= \int d^4x (\frac{1}{2} \phi(-\partial^2+m_\phi^2) \phi+\lambda \phi^4\nonumber\\[4pt]
&+\bar{\psi}\left(\slashed{\partial}+m_\psi\right)\psi-g \phi\bar{\psi}\psi )\,.
\end{align} 
The effective action contains
\begin{align}
\Gamma = &\frac{1}{2}\Gamma^{\phi\phi}\phi\phi-\Gamma^{\bar{\psi}\psi}\bar{\psi}\psi-\Gamma^{\phi\bar{\psi}\psi}\phi\bar{\psi}\psi\nonumber\\[4pt]
&+\frac{1}{24}\Gamma^{\phi\phi\phi\phi}\phi\phi\phi\phi+\frac{1}{4}\Gamma^{\bar{\psi}\bar{\psi}\psi\psi}\bar{\psi}\bar{\psi}\psi\psi\,.
\end{align}
As a master equation we again use the fRG equation,
\begin{verbatim}
fRGEq = {"Prefactor" -> {1/2}, 
<|"type" -> "Regulatordot", 
"indices" -> {i, j}|>, 
<|"type" -> "Propagator", 
"indices" -> {i, j}|>};
\end{verbatim}

\subsubsection{$N_f=1$}

For $N_f=1$ we only have one flavour of fermions and thus only one antifermion/fermion pair in the definition of fields. Furthermore a Yukawa theory also contains a scalar field, which has bosonic statistics.
\begin{verbatim}
fieldsNf1 = <|"bosonic" -> {Phi[p]},
"fermionic" -> 
{{Psibar[p, {d}], Psi[p, {d}]}}|>;
\end{verbatim}
The truncation is given as,
\begin{verbatim}
TruncationfRGNf1 = {{Phi, Phi}, {Psi, Psibar}, 
{Phi, Psi, Psibar}, {Phi, Phi, Phi, Phi},
{Psi, Psi, Psibar, Psibar}};
\end{verbatim}
Thus we can summarize the setup,
\begin{verbatim}
SetupNf1 = <|"MasterEquation" -> fRGEq,
"FieldSpace" -> fieldsNf1, 
"Truncation" -> TruncationfRGNf1|>;
\end{verbatim}

\subsubsection*{\textbf{\textup{Flow of the scalar two-point function}}}

To compute the flow of the scalar two-point function we define the list of derivatives as,
\begin{verbatim}
DerivativeListScalarTwopoint = 
{Phi[-p], Phi[p]};
\end{verbatim}
To get the full diagrams one has to run the command
\begin{verbatim}
fRGDiagramsPhiPhiNf1 = DeriveFunctionalEquation[
SetupNf1, DerivativeListScalarTwopoint, 
"OutputLevel" -> "FullDiagrams"];
\end{verbatim}
The result with superindices $a \simeq (-p)$ and $b\simeq(p)$ is given as,
\begin{align}
\dot{\Gamma}^{\phi_a\phi_b} =
&-\frac{1}{2}R^{\phi\phi}G_{\phi\phi}\Gamma^{\phi\phi\phi_a\phi_b}G_{\phi\phi}\nonumber\\[4pt] &-R^{\bar{\psi}\psi}G_{\psi\bar{\psi}}\Gamma^{\phi_a\bar{\psi}\psi}G_{\psi\bar{\psi}}\Gamma^{\phi_b\bar{\psi}\psi}G_{\psi\bar{\psi}}\nonumber\\[4pt]
&-R^{\bar{\psi}\psi}G_{\psi\bar{\psi}}\Gamma^{\phi_b\bar{\psi}\psi}G_{\psi\bar{\psi}}\Gamma^{\phi_a\bar{\psi}\psi}G_{\psi\bar{\psi}}\,.
\end{align}
The full output of \textit{QMeS} is given in \Cref{app:YukawaNf1}.

\subsubsection*{\textbf{\textup{Flow of the fermionic two-point function}}}

The derivative list for the flow of the fermionic two-point is
\begin{verbatim}
DerivativeListFermionTwopoint = 
{Psibar[-p, {d}], Psi[p, {e}]};
\end{verbatim}
The full diagrams can be obtained with
\begin{verbatim}
fRGDiagramsPsibarPsiNf1 = 
DeriveFunctionalEquation[SetupNf1, 
DerivativeListFermionTwopoint, 
"OutputLevel" -> "FullDiagrams"];
\end{verbatim}
The result with superindices is then given as,
\begin{align}
\dot{\Gamma}^{\bar{\psi}_d\psi_e} =
&-R^{\bar{\psi}\psi}G_{\psi\bar{\psi}}\Gamma^{\bar{\psi}\bar{\psi}_d\psi_e\psi}G_{\psi\bar{\psi}}\nonumber\\[4pt]
&-\frac{1}{2}R^{\phi\phi}G_{\phi\phi}\Gamma^{\phi\bar{\psi}\psi_e}G_{\psi\bar{\psi}}\Gamma^{\phi\bar{\psi}_d\psi}G_{\phi\phi}\nonumber\\[4pt] &-\frac{1}{2}R^{\phi\phi}G_{\phi\phi}\Gamma^{\phi\bar{\psi}_d\psi}G_{\psi\bar{\psi}}\Gamma^{\phi\bar{\psi}\psi_e}G_{\phi\phi}\nonumber\\[4pt]
&-\frac{1}{2}R^{\bar{\psi}\psi}G_{\psi\bar{\psi}}\Gamma^{\phi\bar{\psi}_d\psi}G_{\phi\phi}\Gamma^{\phi\bar{\psi}\psi_e}G_{\psi\bar{\psi}}\nonumber\\[4pt]
&-\frac{1}{2}R^{\bar{\psi}\psi}G_{\psi\bar{\psi}}\Gamma^{\phi\bar{\psi}\psi_e}G_{\phi\phi}\Gamma^{\phi\bar{\psi}_d\psi}G_{\psi\bar{\psi}}\,,
\end{align}
where indices were again dropped. The full output of \textit{QMeS} is given in \Cref{app:YukawaNf1}.

\subsubsection{$N_f=2$}

Since we now want to include two flavours of fermions, we need to implement two antifermion/fermion pairs. For simplicity we call them $(\bar{\psi}_1,\psi_1)$ and $(\bar{\psi}_2,\psi_2)$.
\begin{verbatim}
fieldsNf2 = 
<|"bosonic" -> {Phi[p]},
"fermionic" -> 
{{Psibar1[p, {d}], Psi1[p, {d}]}, 
{Psibar2[p, {d}], Psi2[p, {d}]}}|>;
\end{verbatim}
The truncation is then given by
\begin{verbatim}
TruncationfRGNf2 = {{Phi, Phi}, {Psi1, Psibar1},
{Psi2, Psibar2}, {Phi, Psi1, Psibar1}, 
{Phi, Psi2, Psibar2}, {Phi, Phi, Phi, Phi}, 
{Psi1, Psi1, Psibar1, Psibar1}, 
{Psi2, Psi2, Psibar2, Psibar2}, 
{Psi1, Psi2, Psibar1, Psibar2}};
\end{verbatim}
The setup is then given as,
\begin{verbatim}
SetupNf2 = <|"MasterEquation" -> fRGEq,
"FieldSpace" -> fieldsNf2, 
"Truncation" -> TruncationfRGNf2|>;
\end{verbatim}

\subsubsection*{\textbf{\textup{Flow of the scalar two-point function}}}

As before we define the two scalar field derivatives as,
\begin{verbatim}
DerivativeListScalarTwopoint = 
{Phi[-p], Phi[p]};
\end{verbatim}
To get the full diagrams one has to run the command
\begin{verbatim}
fRGDiagramsPhiPhiNf2 = DeriveFunctionalEquation[
SetupNf2, DerivativeListScalarTwopoint, 
"OutputLevel" -> "FullDiagrams"];
\end{verbatim}
The result in superindex notation with $a \simeq (-p)$ and $b\simeq(p)$ is given as,
\begin{align}
\dot{\Gamma}^{\phi_a\phi_b} =
&-\frac{1}{2}R^{\phi\phi}G_{\phi\phi}\Gamma^{\phi\phi\phi_a\phi_b}G_{\phi\phi}\nonumber\\[4pt] &-R^{\bar{\psi}_1\psi_1}G_{\psi_1\bar{\psi}_1}\Gamma^{\phi_a\bar{\psi}_1\psi_1}G_{\psi_1\bar{\psi}_1}\Gamma^{\phi_b\bar{\psi}_1\psi_1}G_{\psi_1\bar{\psi}_1}\nonumber\\[4pt]
&-R^{\bar{\psi}_1\psi_1}G_{\psi_1\bar{\psi}_1}\Gamma^{\phi_b\bar{\psi}_1\psi_1}G_{\psi_1\bar{\psi}_1}\Gamma^{\phi_a\bar{\psi}_1\psi_1}G_{\psi_1\bar{\psi}_1}\nonumber\\[4pt]
&-R^{\bar{\psi}_2\psi_2}G_{\psi_2\bar{\psi}_2}\Gamma^{\phi_a\bar{\psi}_2\psi_2}G_{\psi_2\bar{\psi}_2}\Gamma^{\phi_b\bar{\psi}_2\psi_2}G_{\psi_2\bar{\psi}_2}\nonumber\\[4pt]
&-R^{\bar{\psi}_2\psi_2}G_{\psi_2\bar{\psi}_2}\Gamma^{\phi_b\bar{\psi}_2\psi_2}G_{\psi_2\bar{\psi}_2}\Gamma^{\phi_a\bar{\psi}_2\psi_2}G_{\psi_2\bar{\psi}_2}\,.
\end{align}
The full output of \textit{QMeS} is given in \Cref{app:YukawaNf2}.

\subsubsection*{\textbf{\textup{Flow of the fermionic two-point function}}}

The derivatives with respect to the first antifermionic and fermionic fields is given as,
\begin{verbatim}
DerivativeListFermion1Twopoint = 
{Psibar1[-p, {d}], Psi1[p, {e}]};
\end{verbatim}
The full diagrams can be obtained with
\begin{verbatim}
fRGDiagramsPsibar1Psi1Nf2 = 
DeriveFunctionalEquation[SetupNf2, 
DerivativeListFermion1Twopoint, 
"OutputLevel" -> "FullDiagrams"];
\end{verbatim}
The result in superindex notation is then
\begin{align}
\dot{\Gamma}^{\bar{\psi}_{1d}\psi_{1e}} =
&-R^{\bar{\psi}_1\psi_1}G_{\psi_1\bar{\psi}_1}\Gamma^{\bar{\psi}_1\bar{\psi}_{1d}\psi_{1e}\psi_1}G_{\psi_1\bar{\psi}_1}\nonumber\\[4pt]
&+R^{\bar{\psi}_2\psi_2}G_{\psi_2\bar{\psi}_2}\Gamma^{\bar{\psi}_{1d}\bar{\psi}_2\psi_{1e}\psi_2}G_{\psi_2\bar{\psi}_2}\nonumber\\[4pt] &-\frac{1}{2}R^{\phi\phi}G_{\phi\phi}\Gamma^{\phi\bar{\psi}_{1d}\psi_1}G_{\psi_1\bar{\psi}_1}\Gamma^{\phi\bar{\psi}_1\psi_{1e}}G_{\phi\phi}\nonumber\\[4pt]
&-\frac{1}{2}R^{\phi\phi}G_{\phi\phi}\Gamma^{\phi\bar{\psi}_1\psi_{1e}}G_{\psi_1\bar{\psi}_1}\Gamma^{\phi\bar{\psi}_{1d}\psi_1}G_{\phi\phi}\nonumber\\[4pt]
&-\frac{1}{2}R^{\bar{\psi}_1\psi_1}G_{\psi_1\bar{\psi}_1}\Gamma^{\phi\bar{\psi}_{1d}\psi_1}G_{\phi\phi}\Gamma^{\phi\bar{\psi}_1\psi_{1e}}G_{\psi_1\bar{\psi}_1}\nonumber\\[4pt]
&-\frac{1}{2}R^{\bar{\psi}_1\psi_1}G_{\psi_1\bar{\psi}_1}\Gamma^{\phi\bar{\psi}_1\psi_{1e}}G_{\phi\phi}\Gamma^{\phi\bar{\psi}_{1d}\psi_1}G_{\psi_1\bar{\psi}_1}\,.
\end{align}
Here one can see how the canonical sorting of fields in vertices is followed by an alphabetical one. The full output of \textit{QMeS} is given in \Cref{app:YukawaNf2}.

\section{Conclusion}
\label{sec:Conclusion}

In this work we have introduced the \textit{Mathematica} package \textit{QMeS-Derivation}. It allows to derive symbolic functional  equations from a master equation (fRG, mSTI, DSE). This includes taking functional derivatives, tracing in field space and a momentum routing for 1-loop diagrams. One of the most notable features is that during this process \textit{QMeS} is able to deal with fermionic signs effectively and consistently. Special emphasis is put on the modular structure of the code which allows for future extensions like for example the extension of the momentum routing to higher loop order diagrams. 

We elucidated the usage of the package by computing different functional equations in $SU(3)$ Yang-Mills and $N_f= 1$ and $N_f=2$ Yukawa theory starting from an action $S$.

\section*{Acknowledgments}

We thank L.~Corell, G.~Eichmann, E.~Grossi, M.~Q.~Huber, F.~Ihssen and J.~Papavassiliou for discussions. This work is also supported by EMMI, and the BMBF grant 05P18VHFCA. It is part of and supported by the DFG Collaborative Research Centre SFB 1225 (ISOQUANT) and the DFG under Germany’s Excellence Strategy EXC - 2181/1 - 390900948 (the Heidelberg Excellence Cluster STRUCTURES).

\appendix

\section{Derivation of Master Equations}
\label{app:Derivation}
In the following we want to outline the basic steps in deriving the \textit{Dyson-Schwinger} and \textit{Functional Renormalisation Group} equations, as well as \textit{(modified) Slavnov-Taylor Identities} that were introduced in \Cref{sec:conventions}.

\subsection{Derivation of the Dyson-Schwinger equation}\label{app:DSEDerivation}
One can derive the \textit{Dyson-Schwinger equation} (DSE) for 1PI Greens functions by taking a total derivative of the integral \labelcref{eq:genFunc}
\begin{align}
0 &=  \int D \phi \frac{\delta}{\delta \phi_a} \exp\left(-S[\phi]+ J^a \phi_a\right)\nonumber\\[4pt]
&=\int D \phi \left(-\frac{\delta S}{\delta \phi_a}+(-1)^{aa}J^a\right) \exp\left(-S[\phi]+ J^a \phi_a\right)\nonumber\\[4pt]
&=\left(-\frac{\delta S}{\delta \phi_a}+(-1)^{aa}J^a\right)_{\phi_b = \frac{\delta}{\delta J^b}} Z[J]\,.
\label{eq:DSEderivation}
\end{align}
When pulling the derivative term the source out of the integral one has to replace the field $\phi$ with a derivative with respect to the source.\\
Using the relation,
\begin{align}
e^{-W[J]}\left(\frac{\delta }{\delta J^a}\right)e^{W[J]} = \frac{\delta W[J]}{\delta J^a}+\frac{\delta }{\delta J^a}\,.
\end{align}
one obtains the DSE for connected Greens functions
\begin{align}
-\frac{\delta S[\phi]}{\delta \phi_a}\Bigg|_{\phi_b = \frac{\delta W[J]}{\delta J^b}+\frac{\delta}{\delta J^b}}+(-1)^{aa}J^a = 0\,.
\end{align}
By rewriting the derivative with respect to $J$ as and using the definition of the propagator $W_{ab} = G_{ab}$,
\begin{align}
\frac{\delta}{\delta J^a} &= \frac{\delta \Phi_b}{\delta J^a} \frac{\delta}{\delta \Phi_b}\nonumber\\[4pt]
&= \frac{\delta}{\delta J^a}\frac{\delta W[J]}{\delta J^b} \frac{\delta}{\delta \Phi_b}\nonumber\\[4pt]
&= G_{ab}\frac{\delta}{\delta \Phi_b}\,,
\end{align} 
one can express the DSE in terms of the effective action
\begin{align}
\frac{\delta \Gamma[\Phi]}{\delta \Phi_a} = \frac{\delta S[\phi]}{\delta \phi_a}\Bigg|_{\phi_b = \Phi_b+G_{bc}\frac{\delta }{\delta \Phi_c}}\,.
\end{align}
The generalized DSE for quantum symmetries can be derived by inserting a generic function $\Psi[\phi]$ in the derivation of equation \labelcref{eq:DSEderivation},
\begin{align}
\frac{1}{Z[J]}\int D\phi \frac{\delta}{\delta \phi_a} \left(\Psi[\phi] \exp{\left(-S[\phi]+J^a\phi_a\right)}\right)\,,
\end{align} 
thus yielding
\begin{align}
\left< \Psi[\phi]\right>\frac{\delta \Gamma[\Phi]}{\delta \Phi_a} = \left< \Psi[\phi]\frac{\delta S[\phi]}{\delta \phi_a}\right>-\left< \frac{\delta \Psi[\phi]}{\delta \phi_a}\right>\,.
\label{eq:generalDSE}
\end{align}

\subsection{Derivation of the fRG equation}
\label{app:fRGderivation}

Since one introduces an (infrared) momentum-regularisation one modifies the Schwinger functional by
\begin{align}
Z[J,R]=e^{W[J,R]} = e^{-\Delta S[\phi,R]}e^{W[J]}\,,
\end{align}
with the so-called regulator insertion,
\begin{align}
\Delta S[\phi,R] = \frac{1}{2}R^{ab}\phi_a\phi_b\,.
\label{eq:reg}
\end{align}
The flow of the generating functional can be written as,
\begin{align}
k \partial_k Z[J,R] &= -\left(k\partial_k \Delta S[\phi,R]\right) Z_k[J,R]\nonumber\\[4pt]
&= -\frac{1}{2}\left(k\partial_k R^{ab}\right)\frac{\delta^2 Z[J,R]}{\delta J^a\delta J^b}\,.
\end{align} 
Using the relation,
\begin{align}
\frac{1}{Z[J,R]}\frac{\delta^2 Z[J,R]}{\delta J^a J^b} = W_{ab}+W_a W_b\,,
\end{align}
the flow equation in terms of the Schwinger functional is
\begin{align}
k\partial_k W = -\frac{1}{2}\left(k\partial_k R^{ab}\right) \left(W_{ab} + W_a W_b\right),
\label{eq:flowSchwingerfunctional}
\end{align}
where $W$ is a function of $J$ and $R$.

One can define the propagator as,
\begin{align}
G_{ac}\left(\Gamma+\Delta S\right)^{cb} = \gamma^b_{\;\,\,a}\,,\nonumber\\[4pt]
\leftrightarrow G_{ac} \left(\Gamma^{cb}+R^{bc}\right) = \gamma^b_{\;\,\,a}\,.
\end{align}
After a Legendre transformation one obtains the effective average action
\begin{align}
\Gamma[\Phi,R] = \underset{J}{\sup}\left(J^a\Phi_a-W[J,R]-\Delta S[\Phi,R]\right).
\label{eq:effavaction}
\end{align}
Again the relations between the fields and sources in terms of the effective average action are given as,
\begin{align}
\frac{\delta(\Gamma[\Phi,R]+\Delta S[\Phi,R])}{\delta \Phi_a} &= (-1)^{aa}J^a\,,\nonumber\\[4pt]
\frac{\delta W[J,R]}{\delta J^a} &= \Phi_a\,.
\label{eq:sources}
\end{align}
By switching to the RG-time $ t = ln(k/ \Lambda)$ with $\partial_t = k\partial_k$ one can write
\begin{align}
\partial_t \Gamma &=-\partial_t W-\partial_t \Delta S\nonumber\\[4pt]
&\;\;\;\;\;-\partial_t J^a\left(\Phi_a -\frac{\delta W}{\delta J^a}\right)\nonumber \\[4pt]
&=\frac{1}{2} \left(\partial_t R^{ab} \right) \left(W_{ab}+W_aW_b\right) -\partial_t \Delta S\nonumber\\[4pt]
&=\frac{1}{2} \left(\partial_t R_k^{ab} \right) W_{ab} +\partial_t \Delta S-\partial_t \Delta S  \nonumber\\[4pt]
&=\frac{1}{2} \dot{R}^{ab} G_{ab}\,.
\end{align}
where $\Gamma\equiv \Gamma[\Phi]$, $W \equiv W[J,R]$, $\Delta S\equiv \Delta S[\Phi,R]$ and we have used equation \labelcref{eq:flowSchwingerfunctional} as well as
\begin{align}
\frac{1}{2}\left(\partial_t R^{ab}\right)\frac{\delta W}{\delta J^a}\frac{\delta W}{\delta J^b} &=\frac{1}{2}\left(\partial_t R^{ab} \right)\Phi_a \Phi_b\nonumber\\[4pt]
&= \partial_t \Delta S\,.
\end{align}
Note that the superfield index notation above implies the summation and thus trace over all fields and integration over space-time.

\subsection{Derivation of the Slavnov-Taylor identity}
\label{app:STIderivation}

The classical Yang-Mills action of non-abelian gauge theories is gauge invariant, but neither the ghost nor the gauge fixed action are, 
\begin{align}
\delta_{gauge}^a e^{-S_A} = \delta_{gauge}^a \left(S_{gf}+S_{gh}\right),
\end{align}
where $\delta_{gauge}^a$ is the generator of a gauge transformation. Additionally it has the form of an operator in the generalized DSE \labelcref{eq:generalDSE} with $\delta/\delta \phi_a \Psi[\phi] = \delta_{gauge}^a$ which means that:
\begin{align}
&\frac{1}{Z[J]} \int D\phi\; \delta_{gauge}^a \left(\exp(-S_A[\phi]+J^a\phi_a)\right)\nonumber\\[4pt]
&= \left< J^a(\delta_{gauge}^a\phi_a)-\delta_{gauge}^a(S_{gf}+S_{gh})\right> =0\,.
\end{align}
Carrying out the expectation value leads to the Slavnov-Taylor identities (STI) of the theory. These identities guarantee the gauge invariance of observables.

Since $\delta_{gauge}^a$ is not a symmetry of the underlying classical theory we would like to find a transformation that is. This is satisfied by the BRST transformation
\begin{align}
\delta_{BRST} \phi_a = \delta \lambda \mathfrak{s}\phi_a\,,
\end{align}
where the infinitesimal parameter $\delta \lambda$ as well as the BRST generator $\mathfrak{s}$ are Grassmannian.

The action is invariant under BRST transformations
\begin{align}
\mathfrak{s} S_A[\phi]=0\,.
\end{align}
Again with $\mathfrak{s}$ as an operator the generalized DSE can be written as,
\begin{align}
&\frac{1}{Z[J]} \int D\phi\; \mathfrak{s} \left(\exp(-S_A[\phi]+J^a\phi_a)\right)=0\,.
\end{align}
Thus the expectation value vanishes,
\begin{align}
(-1)^{aa}\left<J^a \mathfrak{s}\phi_a\right> =0\,.
\label{eq:expBRST}
\end{align}
Note that the prefactor $(-1)^{aa}$ is due to the grassmannian nature of $\mathfrak{s}$.
\begin{align}
\mathfrak{s}J^a\phi_a = (-1)^{aa} J^a \mathfrak{s}\phi_a\,.
\end{align}
Since the BRST transformations of fields are usually quadratic in the fields, it seems as if one looses the algebraic nature of the symmetry on quantum level. To resolve this one may introduce additional source terms $Q^a$ for the BRST transformations of the fields,
\begin{align}
\label{eq:BRSTgenFunc}
Z[J,Q] = \int D\phi \exp(-S_A[\phi]+J^a\phi_a +Q^a \mathfrak{s}\phi_a)\,.
\end{align}
Since $\mathfrak{s}^2 = 0$, this does not change \Cref{eq:expBRST}.

Then one can write the STI takes again algebraic form as
\begin{align}
\left< \mathfrak{s} \phi_a\right> = \frac{1}{Z[J,Q]} \frac{\delta Z[J,Q]}{\delta Q^a}\,.
\end{align}
By Legendre transforming the Schwinger functional $\ln Z[J,Q]$ one obtains the effective action in the presence of source terms for the BRST transformation,
\begin{align}
\Gamma[\Phi,Q] = J^a\Phi_a-\ln Z[J,Q]\,.
\end{align}
We can directly see that
\begin{align}
\left< \mathfrak{s} \phi_a\right> = -\frac{\delta \Gamma[\Phi,Q]}{\delta Q^a} = \frac{1}{Z[J,Q]} \frac{\delta Z[J,Q]}{\delta Q^a}\,.
\end{align}
Rewriting the expectation value \Cref{eq:expBRST} yields the STI
\begin{align}
\frac{\delta \Gamma}{\delta Q^a}\frac{\delta \Gamma}{\delta \Phi_a} =0\,.
\end{align}
Fulfilling this relation guarantees gauge invariance of observables.

\subsection{Derivation of the modified Slavnov-Taylor identity}
\label{app:mSTIderivation}

Due to the presence of the cutoff \labelcref{eq:reg} in the effective average action, gauge and hence BRST symmetry are broken, which means that we need to introduce \textit{modified Slavnov-Taylor identities} (mSTIs) at a non-vanishing momentum scale $k$ that become the usual STIs for $k= 0$. 

Starting from the generating functional with the cutoff term $\Delta S[\phi,R]$ one can derive the mSTIs,
\begin{align}
Z[J,Q] = \int D \phi \exp{\left(-S[\phi]-\Delta S[\phi,R]\right)}\nonumber\\[4pt]
\cdot\exp{\left(J^a\phi_a + Q^a \mathfrak{s}\phi_a\right)}\,.
\end{align}
Note that either the BRST charge or the field itself is of grassmanian nature. Thus the expectation value \Cref{eq:expBRST} changes to
\begin{align}
(-1)^{aa}\left<J^a\mathfrak{s}\phi_a\right> = \left<\mathfrak{s}\Delta S[\phi,R]\right>\,.
\end{align}
After Legendre transforming the momentum scale dependent Schwinger functional one obtains the following relation:
\begin{align}
\left<\mathfrak{s}\phi_a\right> = \frac{\delta W[J,Q,R]}{\delta Q^a}= -\frac{\delta \Gamma[\Phi,Q,R]}{\delta Q^a} \,.
\label{eq:brstSources}
\end{align}
Rewriting the sources $J^a$ in terms of the effective average action \labelcref{eq:sources} one is left with, 
\begin{align}
\frac{\delta \Gamma}{\delta Q^a}\frac{\delta \left(\Gamma+\Delta S\right)}{\delta \Phi_a} = \left<\mathfrak{s}\Delta S[\phi,R]\right>\,.
\end{align}
Moving all terms that contain $\Delta S$ to the right, one can further simplify by using relation \labelcref{eq:brstSources}, 
\begin{align}
&\left\langle\mathfrak{s}\Delta S[\phi,R]\right\rangle-\frac{\delta \Delta S[\Phi,R]}{\delta \Phi_a}\frac{\delta \Gamma[\Phi,Q,R]}{\delta Q^a} \nonumber\\
&=\left\langle\mathfrak{s}\Delta S[\phi,R]\right\rangle +\mathfrak{s}\Delta S[\Phi,R]\,.
\end{align} 
Inserting the cutoff term \labelcref{eq:reg} one arrives at, 
\begin{align}
\left\langle R^{ab}(\mathfrak{s}\phi_a) \phi_b\right\rangle &+ R^{ab}(\mathfrak{s} \Phi_a) \Phi_b\nonumber\\[4pt] 
&= -R^{ab}\frac{\delta}{\delta J^b}\frac{\delta}{\delta Q^a}W[J,Q,R]\nonumber\\[4pt]
&=R^{ab} \frac{\delta }{\delta J^b}\frac{\delta \Gamma[\Phi,Q,R]}{\delta Q^a}\nonumber\\[4pt]
&= R^{ab}G_{bc}\Gamma^{c}_{\;\;Q^a}\,.
\end{align}
Thus the full mSTI reads 
\begin{align}
\frac{\delta \Gamma}{\delta Q^a} \frac{\delta \Gamma}{\delta \Phi_a} = R^{ab}G_{bc}\Gamma^{c}_{\;\;Q^a}\,.
\end{align}
Satisfying the mSTI at each momentum scale $k$ ensures gauge invariance of observables at $k=0$.
\onecolumngrid

\subsection{Summary of derivative rules}
\label{app:DerRules}
The relevant derivative and sign rules that are used in \textit{QMeS} can be summarized as\\

\begin{tabular}{p{0.33\textwidth}p{0.33\textwidth}p{0.33\textwidth}}
	$R^{ab} = (-1)^{ab}R^{ba}\,$, & $G_{ab} = (-1)^{ab} G_{ba}\,$, & 	$\Gamma^{ab} = (-1)^{ab} \Gamma^{ba}\,$,\\[8pt]

	$\frac{\delta}{\delta \Phi_a}\frac{\delta}{\delta \Phi_b} O = (-1)^{ab}\frac{\delta}{\delta \Phi_b}\frac{\delta}{\delta \Phi_a} O\,$,&&\\[8pt]
	
	&&\\[8pt]	
	
	$\frac{\delta}{\delta\phi_a} R^{bc} = 0\,$, &$\frac{\delta}{\delta\phi_a}\phi_b = \delta_{ab}\,$, & $\frac{\delta}{\delta\phi_a} S^{bcd} = 0\,$,\\[8pt]
	
	$\frac{\delta}{\delta\Phi_a} R^{bc} = 0\,$, &$\frac{\delta}{\delta\Phi_a}\Phi_b = \delta_{ab}\,$,&$\frac{\delta}{\delta\Phi_a} S^{bcd} = 0\,$, \\[8pt]
	
	&&\\[8pt]

	$\frac{\delta}{\delta\Phi_a}\Gamma^{b\dots n} = \Gamma^{ab\dots n}\,$,& \multicolumn{2}{l}{$\frac{\delta}{\delta\Phi_a} G_{bc} = (-1)(-1)^{ab}(-1)^{ee}G_{bd}\Gamma^{dae}G_{ec}\,$.}
\end{tabular}

\newpage
\section{Results for the examples}\label{eq:ResultsExamples}

In this section we give the \textit{QMeS} output for the examples in \Cref{sec:example}.

\subsection{YM: Flow of gluon two-point function}
\label{app:fRGappendix}

The full diagrams of the flow of the gluon two-point function with all indices are
\begin{figure}[H]
	\includegraphics[width=1\textwidth]{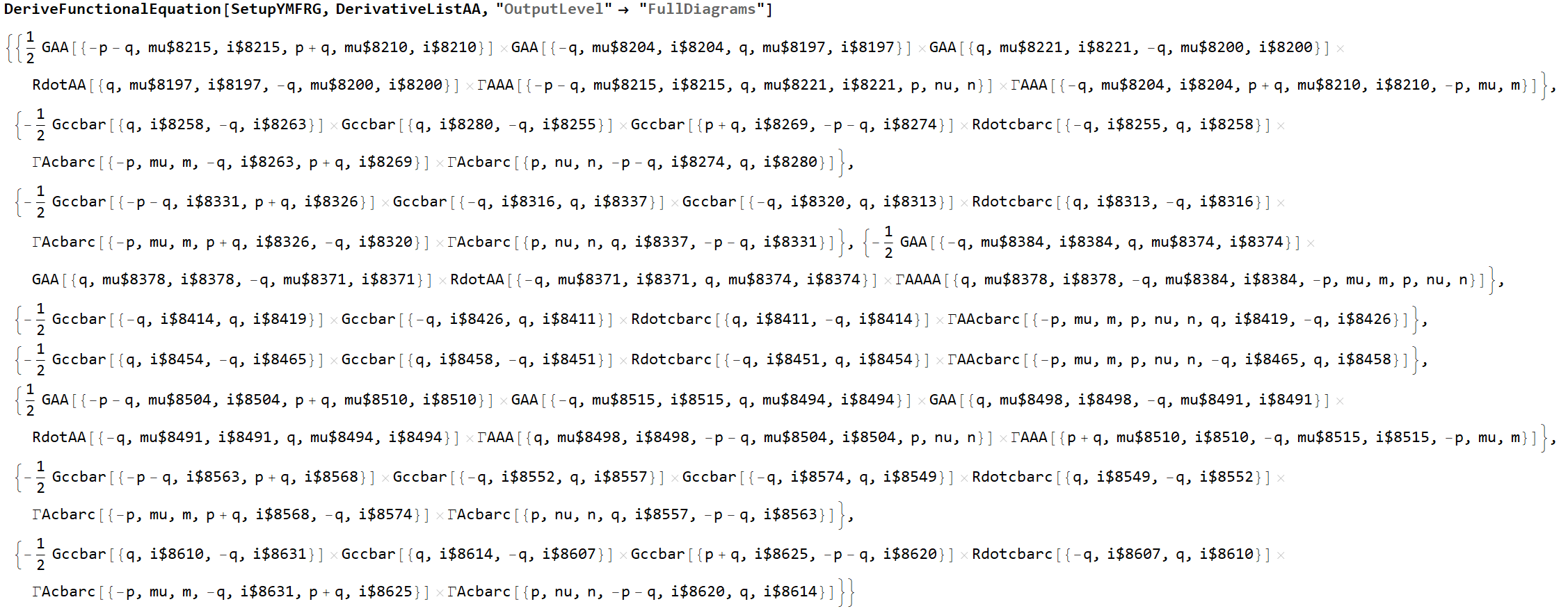}
	\label{fig:frgaa}
\end{figure}
\subsection{YM: mSTI of gluon two-point function}
\label{app:mSTIappendix}
The gluon two-point mSTI is given as:
\begin{figure}[H]
	\includegraphics[width=1\textwidth]{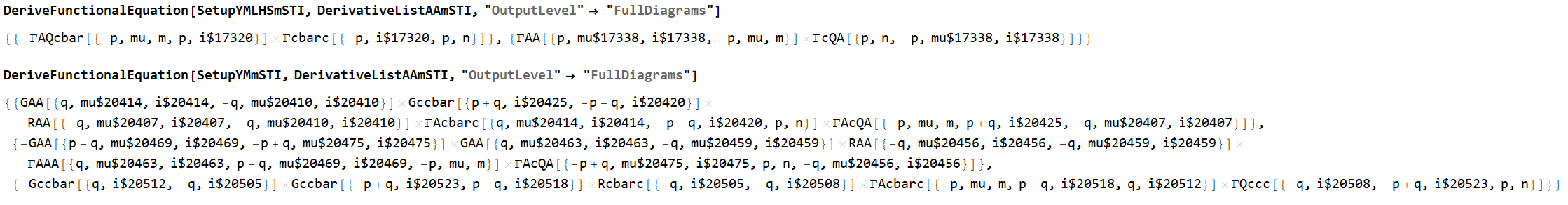}
	\label{fig:mstiaa}
\end{figure}

\subsection{YM: DSE of ghost-gluon vertex}
\label{app:DSEAppendix}

The full ghost-gluon vertex DSE is:
\begin{figure}[H]
	\includegraphics[width=1\textwidth]{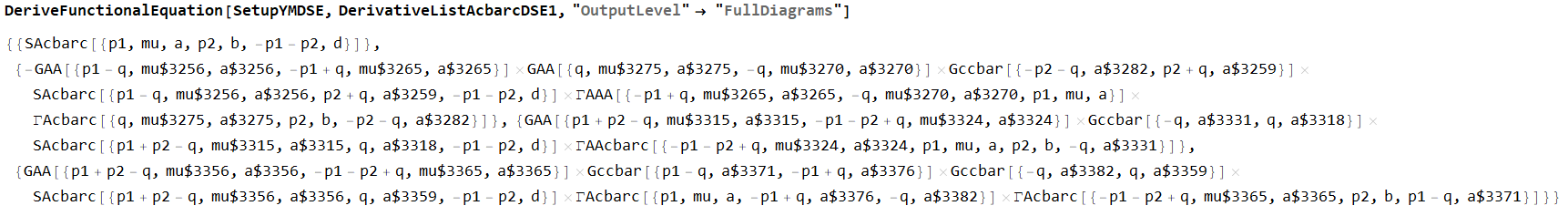}
	\label{fig:dseacbarc}
\end{figure}

\newpage
\subsection{Yukawa $N_f=1$: Flow of two-point functions}
\label{app:YukawaNf1}

The flow of the scalar and fermionic two-point function are:
\begin{figure}[H]
	\includegraphics[width=1\textwidth]{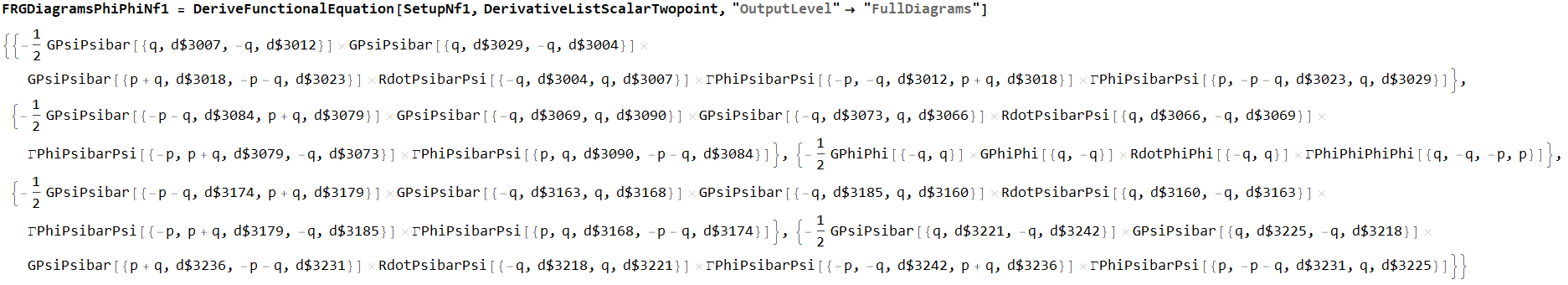}
	\label{fig:frgphiphi1}
\end{figure}
\begin{figure}[H]
	\includegraphics[width=1\textwidth]{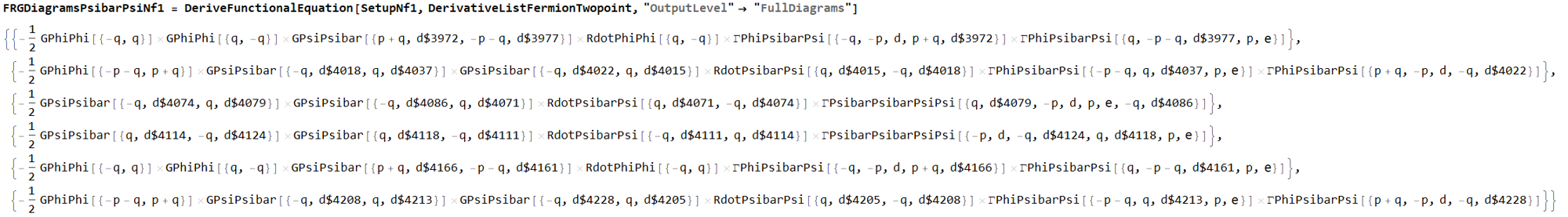}
	\label{fig:frgpsipsi1}
\end{figure}

\subsection{Yukawa $N_f=2$: Flow of two-point functions}
\label{app:YukawaNf2}
The flow of the scalar and fermionic two-point function are:
\begin{figure}[H]
	\includegraphics[width=1\textwidth]{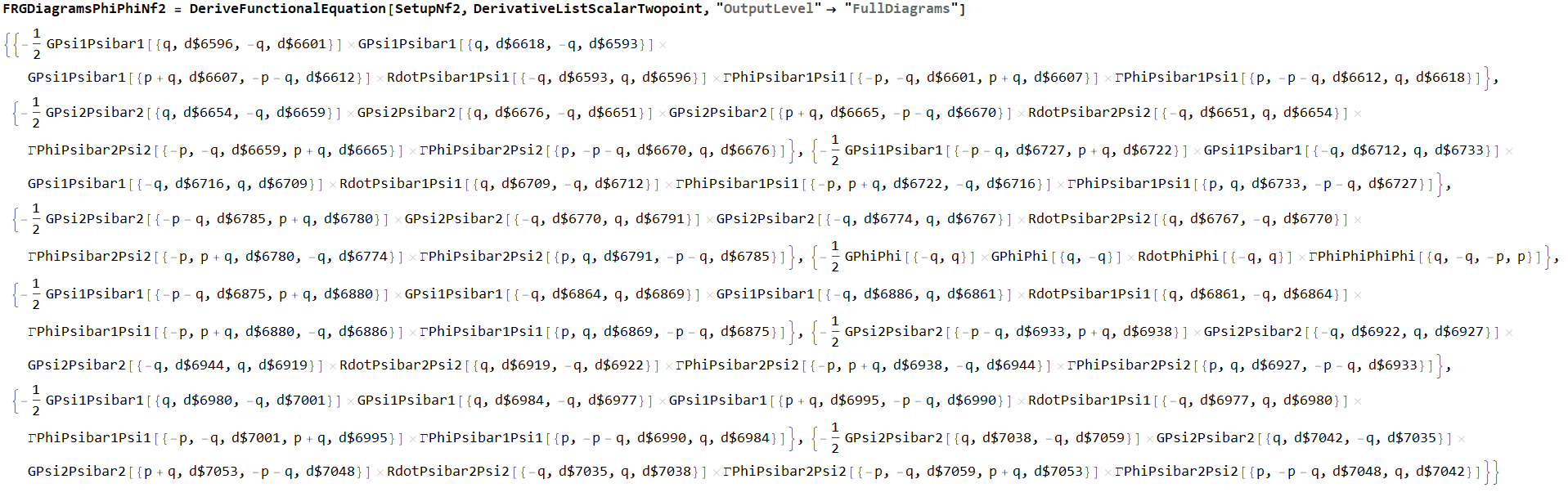}
	\label{fig:frgphiphi2}
\end{figure}
\begin{figure}[H]
	\includegraphics[width=1\textwidth]{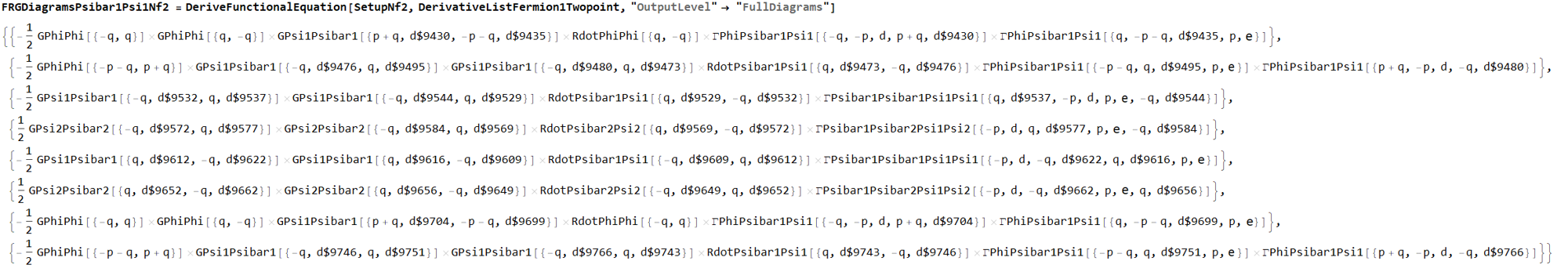}
	\label{fig:frgpsipsi2}
\end{figure}
\newpage
\twocolumngrid

\bibliography{bib_master}

\begin{thebibliography}{53}%
\makeatletter
\providecommand \@ifxundefined [1]{%
 \@ifx{#1\undefined}
}%
\providecommand \@ifnum [1]{%
 \ifnum #1\expandafter \@firstoftwo
 \else \expandafter \@secondoftwo
 \fi
}%
\providecommand \@ifx [1]{%
 \ifx #1\expandafter \@firstoftwo
 \else \expandafter \@secondoftwo
 \fi
}%
\providecommand \natexlab [1]{#1}%
\providecommand \enquote  [1]{``#1''}%
\providecommand \bibnamefont  [1]{#1}%
\providecommand \bibfnamefont [1]{#1}%
\providecommand \citenamefont [1]{#1}%
\providecommand \href@noop [0]{\@secondoftwo}%
\providecommand \href [0]{\begingroup \@sanitize@url \@href}%
\providecommand \@href[1]{\@@startlink{#1}\@@href}%
\providecommand \@@href[1]{\endgroup#1\@@endlink}%
\providecommand \@sanitize@url [0]{\catcode `\\12\catcode `\$12\catcode
  `\&12\catcode `\#12\catcode `\^12\catcode `\_12\catcode `\%12\relax}%
\providecommand \@@startlink[1]{}%
\providecommand \@@endlink[0]{}%
\providecommand \url  [0]{\begingroup\@sanitize@url \@url }%
\providecommand \@url [1]{\endgroup\@href {#1}{\urlprefix }}%
\providecommand \urlprefix  [0]{URL }%
\providecommand \Eprint [0]{\href }%
\providecommand \doibase [0]{https://doi.org/}%
\providecommand \selectlanguage [0]{\@gobble}%
\providecommand \bibinfo  [0]{\@secondoftwo}%
\providecommand \bibfield  [0]{\@secondoftwo}%
\providecommand \translation [1]{[#1]}%
\providecommand \BibitemOpen [0]{}%
\providecommand \bibitemStop [0]{}%
\providecommand \bibitemNoStop [0]{.\EOS\space}%
\providecommand \EOS [0]{\spacefactor3000\relax}%
\providecommand \BibitemShut  [1]{\csname bibitem#1\endcsname}%
\let\auto@bib@innerbib\@empty
\bibitem [{\citenamefont {Berges}\ \emph {et~al.}(2002)\citenamefont {Berges},
  \citenamefont {Tetradis},\ and\ \citenamefont {Wetterich}}]{Berges:2000ew}%
  \BibitemOpen
  \bibfield  {author} {\bibinfo {author} {\bibfnamefont {J.}~\bibnamefont
  {Berges}}, \bibinfo {author} {\bibfnamefont {N.}~\bibnamefont {Tetradis}},\
  and\ \bibinfo {author} {\bibfnamefont {C.}~\bibnamefont {Wetterich}},\
  }\bibfield  {title} {\bibinfo {title} {{Non-perturbative renormalization flow
  in quantum field theory and statistical physics}},\ }\href
  {https://doi.org/10.1016/S0370-1573(01)00098-9} {\bibfield  {journal}
  {\bibinfo  {journal} {Phys. Rept.}\ }\textbf {\bibinfo {volume} {363}},\
  \bibinfo {pages} {223} (\bibinfo {year} {2002})},\ \Eprint
  {https://arxiv.org/abs/hep-ph/0005122} {arXiv:hep-ph/0005122} \BibitemShut
  {NoStop}%
\bibitem [{\citenamefont {Polonyi}(2003)}]{Polonyi:2001se}%
  \BibitemOpen
  \bibfield  {author} {\bibinfo {author} {\bibfnamefont {J.}~\bibnamefont
  {Polonyi}},\ }\bibfield  {title} {\bibinfo {title} {{Lectures on the
  functional renormalization group method}},\ }\href
  {https://doi.org/10.2478/BF02475552} {\bibfield  {journal} {\bibinfo
  {journal} {Central Eur.J.Phys.}\ }\textbf {\bibinfo {volume} {1}},\ \bibinfo
  {pages} {1} (\bibinfo {year} {2003})},\ \Eprint
  {https://arxiv.org/abs/hep-th/0110026} {arXiv:hep-th/0110026 [hep-th]}
  \BibitemShut {NoStop}%
\bibitem [{\citenamefont {Delamotte}\ \emph {et~al.}(2004)\citenamefont
  {Delamotte}, \citenamefont {Mouhanna},\ and\ \citenamefont
  {Tissier}}]{Delamotte:2003dw}%
  \BibitemOpen
  \bibfield  {author} {\bibinfo {author} {\bibfnamefont {B.}~\bibnamefont
  {Delamotte}}, \bibinfo {author} {\bibfnamefont {D.}~\bibnamefont
  {Mouhanna}},\ and\ \bibinfo {author} {\bibfnamefont {M.}~\bibnamefont
  {Tissier}},\ }\bibfield  {title} {\bibinfo {title} {{Nonperturbative
  renormalization group approach to frustrated magnets}},\ }\href
  {https://doi.org/10.1103/PhysRevB.69.134413} {\bibfield  {journal} {\bibinfo
  {journal} {Phys.Rev.}\ }\textbf {\bibinfo {volume} {B69}},\ \bibinfo {pages}
  {134413} (\bibinfo {year} {2004})},\ \Eprint
  {https://arxiv.org/abs/cond-mat/0309101} {arXiv:cond-mat/0309101 [cond-mat]}
  \BibitemShut {NoStop}%
\bibitem [{\citenamefont {Pawlowski}(2007)}]{Pawlowski:2005xe}%
  \BibitemOpen
  \bibfield  {author} {\bibinfo {author} {\bibfnamefont {J.~M.}\ \bibnamefont
  {Pawlowski}},\ }\bibfield  {title} {\bibinfo {title} {{Aspects of the
  functional renormalisation group}},\ }\href
  {https://doi.org/10.1016/j.aop.2007.01.007} {\bibfield  {journal} {\bibinfo
  {journal} {Annals Phys.}\ }\textbf {\bibinfo {volume} {322}},\ \bibinfo
  {pages} {2831} (\bibinfo {year} {2007})},\ \Eprint
  {https://arxiv.org/abs/hep-th/0512261} {arXiv:hep-th/0512261 [hep-th]}
  \BibitemShut {NoStop}%
\bibitem [{\citenamefont {Schaefer}\ and\ \citenamefont
  {Wambach}(2008)}]{Schaefer:2006sr}%
  \BibitemOpen
  \bibfield  {author} {\bibinfo {author} {\bibfnamefont {B.-J.}\ \bibnamefont
  {Schaefer}}\ and\ \bibinfo {author} {\bibfnamefont {J.}~\bibnamefont
  {Wambach}},\ }\bibfield  {title} {\bibinfo {title} {{Renormalization group
  approach towards the QCD phase diagram}},\ }\href
  {https://doi.org/10.1134/S1063779608070083} {\bibfield  {journal} {\bibinfo
  {journal} {Phys.Part.Nucl.}\ }\textbf {\bibinfo {volume} {39}},\ \bibinfo
  {pages} {1025} (\bibinfo {year} {2008})},\ \Eprint
  {https://arxiv.org/abs/hep-ph/0611191} {arXiv:hep-ph/0611191 [hep-ph]}
  \BibitemShut {NoStop}%
\bibitem [{\citenamefont {Gies}(2012)}]{Gies:2006wv}%
  \BibitemOpen
  \bibfield  {author} {\bibinfo {author} {\bibfnamefont {H.}~\bibnamefont
  {Gies}},\ }\bibfield  {title} {\bibinfo {title} {{Introduction to the
  functional RG and applications to gauge theories}},\ }\href
  {https://doi.org/10.1007/978-3-642-27320-9_6} {\bibfield  {journal} {\bibinfo
   {journal} {Lect.Notes Phys.}\ }\textbf {\bibinfo {volume} {852}},\ \bibinfo
  {pages} {287} (\bibinfo {year} {2012})},\ \Eprint
  {https://arxiv.org/abs/hep-ph/0611146} {arXiv:hep-ph/0611146 [hep-ph]}
  \BibitemShut {NoStop}%
\bibitem [{\citenamefont {Delamotte}(2012)}]{Delamotte:2007pf}%
  \BibitemOpen
  \bibfield  {author} {\bibinfo {author} {\bibfnamefont {B.}~\bibnamefont
  {Delamotte}},\ }\bibfield  {title} {\bibinfo {title} {{An Introduction to the
  nonperturbative renormalization group}},\ }\href
  {https://doi.org/10.1007/978-3-642-27320-9_2} {\bibfield  {journal} {\bibinfo
   {journal} {Lect. Notes Phys.}\ }\textbf {\bibinfo {volume} {852}},\ \bibinfo
  {pages} {49} (\bibinfo {year} {2012})},\ \Eprint
  {https://arxiv.org/abs/cond-mat/0702365} {arXiv:cond-mat/0702365
  [cond-mat.stat-mech]} \BibitemShut {NoStop}%
\bibitem [{\citenamefont {Igarashi}\ \emph {et~al.}(2010)\citenamefont
  {Igarashi}, \citenamefont {Itoh},\ and\ \citenamefont
  {Sonoda}}]{Igarashi:2009tj}%
  \BibitemOpen
  \bibfield  {author} {\bibinfo {author} {\bibfnamefont {Y.}~\bibnamefont
  {Igarashi}}, \bibinfo {author} {\bibfnamefont {K.}~\bibnamefont {Itoh}},\
  and\ \bibinfo {author} {\bibfnamefont {H.}~\bibnamefont {Sonoda}},\
  }\bibfield  {title} {\bibinfo {title} {{Realization of Symmetry in the ERG
  Approach to Quantum Field Theory}},\ }\href
  {https://doi.org/10.1143/PTPS.181.1} {\bibfield  {journal} {\bibinfo
  {journal} {Prog. Theor. Phys. Suppl.}\ }\textbf {\bibinfo {volume} {181}},\
  \bibinfo {pages} {1} (\bibinfo {year} {2010})},\ \Eprint
  {https://arxiv.org/abs/0909.0327} {arXiv:0909.0327 [hep-th]} \BibitemShut
  {NoStop}%
\bibitem [{\citenamefont {Rosten}(2012)}]{Rosten:2010vm}%
  \BibitemOpen
  \bibfield  {author} {\bibinfo {author} {\bibfnamefont {O.~J.}\ \bibnamefont
  {Rosten}},\ }\bibfield  {title} {\bibinfo {title} {{Fundamentals of the Exact
  Renormalization Group}},\ }\href
  {https://doi.org/10.1016/j.physrep.2011.12.003} {\bibfield  {journal}
  {\bibinfo  {journal} {Phys. Rept.}\ }\textbf {\bibinfo {volume} {511}},\
  \bibinfo {pages} {177} (\bibinfo {year} {2012})},\ \Eprint
  {https://arxiv.org/abs/1003.1366} {arXiv:1003.1366 [hep-th]} \BibitemShut
  {NoStop}%
\bibitem [{\citenamefont {Kopietz}\ \emph {et~al.}(2010)\citenamefont
  {Kopietz}, \citenamefont {Bartosch},\ and\ \citenamefont
  {Sch\"utz}}]{Kopietz:2010zz}%
  \BibitemOpen
  \bibfield  {author} {\bibinfo {author} {\bibfnamefont {P.}~\bibnamefont
  {Kopietz}}, \bibinfo {author} {\bibfnamefont {L.}~\bibnamefont {Bartosch}},\
  and\ \bibinfo {author} {\bibfnamefont {F.}~\bibnamefont {Sch\"utz}},\ }\href
  {https://doi.org/10.1007/978-3-642-05094-7} {\emph {\bibinfo {title}
  {{Introduction to the functional renormalization group}}}},\ Vol.\ \bibinfo
  {volume} {798}\ (\bibinfo {year} {2010})\BibitemShut {NoStop}%
\bibitem [{\citenamefont {Braun}(2012)}]{Braun:2011pp}%
  \BibitemOpen
  \bibfield  {author} {\bibinfo {author} {\bibfnamefont {J.}~\bibnamefont
  {Braun}},\ }\bibfield  {title} {\bibinfo {title} {{Fermion Interactions and
  Universal Behavior in Strongly Interacting Theories}},\ }\href
  {https://doi.org/10.1088/0954-3899/39/3/033001} {\bibfield  {journal}
  {\bibinfo  {journal} {J.Phys.}\ }\textbf {\bibinfo {volume} {G39}},\ \bibinfo
  {pages} {033001} (\bibinfo {year} {2012})},\ \Eprint
  {https://arxiv.org/abs/1108.4449} {arXiv:1108.4449 [hep-ph]} \BibitemShut
  {NoStop}%
\bibitem [{\citenamefont {Litim}(2011)}]{Litim:2011cp}%
  \BibitemOpen
  \bibfield  {author} {\bibinfo {author} {\bibfnamefont {D.~F.}\ \bibnamefont
  {Litim}},\ }\bibfield  {title} {\bibinfo {title} {{Renormalisation group and
  the Planck scale}},\ }\href {https://doi.org/10.1098/rsta.2011.0103}
  {\bibfield  {journal} {\bibinfo  {journal} {Phil. Trans. Roy. Soc. Lond. A}\
  }\textbf {\bibinfo {volume} {369}},\ \bibinfo {pages} {2759} (\bibinfo {year}
  {2011})},\ \Eprint {https://arxiv.org/abs/1102.4624} {arXiv:1102.4624
  [hep-th]} \BibitemShut {NoStop}%
\bibitem [{\citenamefont {Metzner}\ \emph {et~al.}(2012)\citenamefont
  {Metzner}, \citenamefont {Salmhofer}, \citenamefont {Honerkamp},
  \citenamefont {Meden},\ and\ \citenamefont {Schönhammer}}]{Metzner_2012}%
  \BibitemOpen
  \bibfield  {author} {\bibinfo {author} {\bibfnamefont {W.}~\bibnamefont
  {Metzner}}, \bibinfo {author} {\bibfnamefont {M.}~\bibnamefont {Salmhofer}},
  \bibinfo {author} {\bibfnamefont {C.}~\bibnamefont {Honerkamp}}, \bibinfo
  {author} {\bibfnamefont {V.}~\bibnamefont {Meden}},\ and\ \bibinfo {author}
  {\bibfnamefont {K.}~\bibnamefont {Schönhammer}},\ }\bibfield  {title}
  {\bibinfo {title} {Functional renormalization group approach to correlated
  fermion systems},\ }\href {https://doi.org/10.1103/revmodphys.84.299}
  {\bibfield  {journal} {\bibinfo  {journal} {Reviews of Modern Physics}\
  }\textbf {\bibinfo {volume} {84}},\ \bibinfo {pages} {299–352} (\bibinfo
  {year} {2012})}\BibitemShut {NoStop}%
\bibitem [{\citenamefont {Salmhofer}(2019)}]{Salmhofer:2018sgo}%
  \BibitemOpen
  \bibfield  {author} {\bibinfo {author} {\bibfnamefont {M.}~\bibnamefont
  {Salmhofer}},\ }\bibfield  {title} {\bibinfo {title} {{Renormalization in
  condensed matter: Fermionic systems \textendash{} from mathematics to
  materials}},\ }\href {https://doi.org/10.1016/j.nuclphysb.2018.07.004}
  {\bibfield  {journal} {\bibinfo  {journal} {Nucl. Phys. B}\ }\textbf
  {\bibinfo {volume} {941}},\ \bibinfo {pages} {868} (\bibinfo {year}
  {2019})},\ \Eprint {https://arxiv.org/abs/1807.01766} {arXiv:1807.01766
  [cond-mat.str-el]} \BibitemShut {NoStop}%
\bibitem [{\citenamefont {Eichhorn}(2019)}]{Eichhorn:2018yfc}%
  \BibitemOpen
  \bibfield  {author} {\bibinfo {author} {\bibfnamefont {A.}~\bibnamefont
  {Eichhorn}},\ }\bibfield  {title} {\bibinfo {title} {{An asymptotically safe
  guide to quantum gravity and matter}},\ }\href
  {https://doi.org/10.3389/fspas.2018.00047} {\bibfield  {journal} {\bibinfo
  {journal} {Front. Astron. Space Sci.}\ }\textbf {\bibinfo {volume} {5}},\
  \bibinfo {pages} {47} (\bibinfo {year} {2019})},\ \Eprint
  {https://arxiv.org/abs/1810.07615} {arXiv:1810.07615 [hep-th]} \BibitemShut
  {NoStop}%
\bibitem [{\citenamefont {Reuter}\ and\ \citenamefont
  {Saueressig}(2019)}]{Reuter:2019byg}%
  \BibitemOpen
  \bibfield  {author} {\bibinfo {author} {\bibfnamefont {M.}~\bibnamefont
  {Reuter}}\ and\ \bibinfo {author} {\bibfnamefont {F.}~\bibnamefont
  {Saueressig}},\ }\href@noop {} {\emph {\bibinfo {title} {{Quantum Gravity and
  the Functional Renormalization Group}: {The Road towards Asymptotic
  Safety}}}}\ (\bibinfo  {publisher} {Cambridge University Press},\ \bibinfo
  {year} {2019})\BibitemShut {NoStop}%
\bibitem [{\citenamefont {Bonanno}\ \emph {et~al.}(2020)\citenamefont
  {Bonanno}, \citenamefont {Eichhorn}, \citenamefont {Gies}, \citenamefont
  {Pawlowski}, \citenamefont {Percacci}, \citenamefont {Reuter}, \citenamefont
  {Saueressig},\ and\ \citenamefont {Vacca}}]{Bonanno:2020bil}%
  \BibitemOpen
  \bibfield  {author} {\bibinfo {author} {\bibfnamefont {A.}~\bibnamefont
  {Bonanno}}, \bibinfo {author} {\bibfnamefont {A.}~\bibnamefont {Eichhorn}},
  \bibinfo {author} {\bibfnamefont {H.}~\bibnamefont {Gies}}, \bibinfo {author}
  {\bibfnamefont {J.~M.}\ \bibnamefont {Pawlowski}}, \bibinfo {author}
  {\bibfnamefont {R.}~\bibnamefont {Percacci}}, \bibinfo {author}
  {\bibfnamefont {M.}~\bibnamefont {Reuter}}, \bibinfo {author} {\bibfnamefont
  {F.}~\bibnamefont {Saueressig}},\ and\ \bibinfo {author} {\bibfnamefont
  {G.~P.}\ \bibnamefont {Vacca}},\ }\bibfield  {title} {\bibinfo {title}
  {{Critical reflections on asymptotically safe gravity}},\ }\href
  {https://doi.org/10.3389/fphy.2020.00269} {\bibfield  {journal} {\bibinfo
  {journal} {Front. in Phys.}\ }\textbf {\bibinfo {volume} {8}},\ \bibinfo
  {pages} {269} (\bibinfo {year} {2020})},\ \Eprint
  {https://arxiv.org/abs/2004.06810} {arXiv:2004.06810 [gr-qc]} \BibitemShut
  {NoStop}%
\bibitem [{\citenamefont {Dupuis}\ \emph {et~al.}(2021)\citenamefont {Dupuis},
  \citenamefont {Canet}, \citenamefont {Eichhorn}, \citenamefont {Metzner},
  \citenamefont {Pawlowski}, \citenamefont {Tissier},\ and\ \citenamefont
  {Wschebor}}]{Dupuis:2020fhh}%
  \BibitemOpen
  \bibfield  {author} {\bibinfo {author} {\bibfnamefont {N.}~\bibnamefont
  {Dupuis}}, \bibinfo {author} {\bibfnamefont {L.}~\bibnamefont {Canet}},
  \bibinfo {author} {\bibfnamefont {A.}~\bibnamefont {Eichhorn}}, \bibinfo
  {author} {\bibfnamefont {W.}~\bibnamefont {Metzner}}, \bibinfo {author}
  {\bibfnamefont {J.}~\bibnamefont {Pawlowski}}, \bibinfo {author}
  {\bibfnamefont {M.}~\bibnamefont {Tissier}},\ and\ \bibinfo {author}
  {\bibfnamefont {N.}~\bibnamefont {Wschebor}},\ }\bibfield  {title} {\bibinfo
  {title} {The nonperturbative functional renormalization group and its
  applications},\ }\bibfield  {journal} {\bibinfo  {journal} {Physics Reports}\
  }\href {https://doi.org/https://doi.org/10.1016/j.physrep.2021.01.001}
  {https://doi.org/10.1016/j.physrep.2021.01.001} (\bibinfo {year}
  {2021})\BibitemShut {NoStop}%
\bibitem [{\citenamefont {Pawlowski}\ and\ \citenamefont
  {Reichert}(2020)}]{Pawlowski:2020qer}%
  \BibitemOpen
  \bibfield  {author} {\bibinfo {author} {\bibfnamefont {J.~M.}\ \bibnamefont
  {Pawlowski}}\ and\ \bibinfo {author} {\bibfnamefont {M.}~\bibnamefont
  {Reichert}},\ }\href@noop {} {\bibinfo {title} {Quantum gravity: a
  fluctuating point of view}} (\bibinfo {year} {2020}),\ \Eprint
  {https://arxiv.org/abs/2007.10353} {arXiv:2007.10353 [hep-th]} \BibitemShut
  {NoStop}%
\bibitem [{\citenamefont {Roberts}\ and\ \citenamefont
  {Schmidt}(2000)}]{Roberts:2000aa}%
  \BibitemOpen
  \bibfield  {author} {\bibinfo {author} {\bibfnamefont {C.~D.}\ \bibnamefont
  {Roberts}}\ and\ \bibinfo {author} {\bibfnamefont {S.~M.}\ \bibnamefont
  {Schmidt}},\ }\bibfield  {title} {\bibinfo {title} {{Dyson-Schwinger
  equations: Density, temperature and continuum strong QCD}},\ }\href@noop {}
  {\bibfield  {journal} {\bibinfo  {journal} {Prog. Part. Nucl. Phys.}\
  }\textbf {\bibinfo {volume} {45}},\ \bibinfo {pages} {S1} (\bibinfo {year}
  {2000})},\ \Eprint {https://arxiv.org/abs/nucl-th/0005064}
  {arXiv:nucl-th/0005064} \BibitemShut {NoStop}%
\bibitem [{\citenamefont {Alkofer}\ and\ \citenamefont {von
  Smekal}(2001)}]{Alkofer:2000wg}%
  \BibitemOpen
  \bibfield  {author} {\bibinfo {author} {\bibfnamefont {R.}~\bibnamefont
  {Alkofer}}\ and\ \bibinfo {author} {\bibfnamefont {L.}~\bibnamefont {von
  Smekal}},\ }\bibfield  {title} {\bibinfo {title} {{The infrared behavior of
  QCD Green's functions: Confinement, dynamical symmetry breaking, and hadrons
  as relativistic bound states}},\ }\href
  {https://doi.org/10.1016/S0370-1573(01)00010-2} {\bibfield  {journal}
  {\bibinfo  {journal} {Phys. Rept.}\ }\textbf {\bibinfo {volume} {353}},\
  \bibinfo {pages} {281} (\bibinfo {year} {2001})},\ \Eprint
  {https://arxiv.org/abs/hep-ph/0007355} {arXiv:hep-ph/0007355} \BibitemShut
  {NoStop}%
\bibitem [{\citenamefont {Fischer}(2006)}]{Fischer:2006ub}%
  \BibitemOpen
  \bibfield  {author} {\bibinfo {author} {\bibfnamefont {C.~S.}\ \bibnamefont
  {Fischer}},\ }\bibfield  {title} {\bibinfo {title} {{Infrared properties of
  QCD from Dyson-Schwinger equations}},\ }\href
  {https://doi.org/10.1088/0954-3899/32/8/R02} {\bibfield  {journal} {\bibinfo
  {journal} {J.Phys.G}\ }\textbf {\bibinfo {volume} {G32}},\ \bibinfo {pages}
  {R253} (\bibinfo {year} {2006})},\ \Eprint
  {https://arxiv.org/abs/hep-ph/0605173} {arXiv:hep-ph/0605173 [hep-ph]}
  \BibitemShut {NoStop}%
\bibitem [{\citenamefont {Binosi}\ and\ \citenamefont
  {Papavassiliou}(2009)}]{Binosi:2009qm}%
  \BibitemOpen
  \bibfield  {author} {\bibinfo {author} {\bibfnamefont {D.}~\bibnamefont
  {Binosi}}\ and\ \bibinfo {author} {\bibfnamefont {J.}~\bibnamefont
  {Papavassiliou}},\ }\bibfield  {title} {\bibinfo {title} {{Pinch Technique:
  Theory and Applications}},\ }\href
  {https://doi.org/10.1016/j.physrep.2009.05.001} {\bibfield  {journal}
  {\bibinfo  {journal} {Phys. Rept.}\ }\textbf {\bibinfo {volume} {479}},\
  \bibinfo {pages} {1} (\bibinfo {year} {2009})},\ \Eprint
  {https://arxiv.org/abs/0909.2536} {arXiv:0909.2536 [hep-ph]} \BibitemShut
  {NoStop}%
\bibitem [{\citenamefont {Maas}(2013)}]{Maas:2011se}%
  \BibitemOpen
  \bibfield  {author} {\bibinfo {author} {\bibfnamefont {A.}~\bibnamefont
  {Maas}},\ }\bibfield  {title} {\bibinfo {title} {{Describing gauge bosons at
  zero and finite temperature}},\ }\href
  {https://doi.org/10.1016/j.physrep.2012.11.002} {\bibfield  {journal}
  {\bibinfo  {journal} {Phys.Rept.}\ }\textbf {\bibinfo {volume} {524}},\
  \bibinfo {pages} {203} (\bibinfo {year} {2013})},\ \Eprint
  {https://arxiv.org/abs/1106.3942} {arXiv:1106.3942 [hep-ph]} \BibitemShut
  {NoStop}%
\bibitem [{\citenamefont {Boucaud}\ \emph {et~al.}(2012)\citenamefont
  {Boucaud}, \citenamefont {Leroy}, \citenamefont {Yaouanc}, \citenamefont
  {Micheli}, \citenamefont {Pene},\ and\ \citenamefont
  {Rodriguez-Quintero}}]{Boucaud:2011ug}%
  \BibitemOpen
  \bibfield  {author} {\bibinfo {author} {\bibfnamefont {P.}~\bibnamefont
  {Boucaud}}, \bibinfo {author} {\bibfnamefont {J.~P.}\ \bibnamefont {Leroy}},
  \bibinfo {author} {\bibfnamefont {A.~L.}\ \bibnamefont {Yaouanc}}, \bibinfo
  {author} {\bibfnamefont {J.}~\bibnamefont {Micheli}}, \bibinfo {author}
  {\bibfnamefont {O.}~\bibnamefont {Pene}},\ and\ \bibinfo {author}
  {\bibfnamefont {J.}~\bibnamefont {Rodriguez-Quintero}},\ }\bibfield  {title}
  {\bibinfo {title} {{The Infrared Behaviour of the Pure Yang-Mills Green
  Functions}},\ }\href {https://doi.org/10.1007/s00601-011-0301-2} {\bibfield
  {journal} {\bibinfo  {journal} {Few Body Syst.}\ }\textbf {\bibinfo {volume}
  {53}},\ \bibinfo {pages} {387} (\bibinfo {year} {2012})},\ \Eprint
  {https://arxiv.org/abs/1109.1936} {arXiv:1109.1936 [hep-ph]} \BibitemShut
  {NoStop}%
\bibitem [{\citenamefont {Aguilar}\ \emph {et~al.}(2016)\citenamefont
  {Aguilar}, \citenamefont {Binosi},\ and\ \citenamefont
  {Papavassiliou}}]{Aguilar:2015bud}%
  \BibitemOpen
  \bibfield  {author} {\bibinfo {author} {\bibfnamefont {A.~C.}\ \bibnamefont
  {Aguilar}}, \bibinfo {author} {\bibfnamefont {D.}~\bibnamefont {Binosi}},\
  and\ \bibinfo {author} {\bibfnamefont {J.}~\bibnamefont {Papavassiliou}},\
  }\bibfield  {title} {\bibinfo {title} {{The Gluon Mass Generation Mechanism:
  A Concise Primer}},\ }\href {https://doi.org/10.1007/s11467-015-0517-6}
  {\bibfield  {journal} {\bibinfo  {journal} {Front. Phys. (Beijing)}\ }\textbf
  {\bibinfo {volume} {11}},\ \bibinfo {pages} {111203} (\bibinfo {year}
  {2016})},\ \Eprint {https://arxiv.org/abs/1511.08361} {arXiv:1511.08361
  [hep-ph]} \BibitemShut {NoStop}%
\bibitem [{\citenamefont {Eichmann}\ \emph {et~al.}(2016)\citenamefont
  {Eichmann}, \citenamefont {Sanchis-Alepuz}, \citenamefont {Williams},
  \citenamefont {Alkofer},\ and\ \citenamefont {Fischer}}]{Eichmann:2016yit}%
  \BibitemOpen
  \bibfield  {author} {\bibinfo {author} {\bibfnamefont {G.}~\bibnamefont
  {Eichmann}}, \bibinfo {author} {\bibfnamefont {H.}~\bibnamefont
  {Sanchis-Alepuz}}, \bibinfo {author} {\bibfnamefont {R.}~\bibnamefont
  {Williams}}, \bibinfo {author} {\bibfnamefont {R.}~\bibnamefont {Alkofer}},\
  and\ \bibinfo {author} {\bibfnamefont {C.~S.}\ \bibnamefont {Fischer}},\
  }\bibfield  {title} {\bibinfo {title} {{Baryons as relativistic three-quark
  bound states}},\ }\href {https://doi.org/10.1016/j.ppnp.2016.07.001}
  {\bibfield  {journal} {\bibinfo  {journal} {Prog. Part. Nucl. Phys.}\
  }\textbf {\bibinfo {volume} {91}},\ \bibinfo {pages} {1} (\bibinfo {year}
  {2016})},\ \Eprint {https://arxiv.org/abs/1606.09602} {arXiv:1606.09602
  [hep-ph]} \BibitemShut {NoStop}%
\bibitem [{\citenamefont {Sanchis-Alepuz}\ and\ \citenamefont
  {Williams}(2018)}]{Sanchis-Alepuz:2017jjd}%
  \BibitemOpen
  \bibfield  {author} {\bibinfo {author} {\bibfnamefont {H.}~\bibnamefont
  {Sanchis-Alepuz}}\ and\ \bibinfo {author} {\bibfnamefont {R.}~\bibnamefont
  {Williams}},\ }\bibfield  {title} {\bibinfo {title} {{Recent developments in
  bound-state calculations using the Dyson\textendash{}Schwinger and
  Bethe\textendash{}Salpeter equations}},\ }\href
  {https://doi.org/10.1016/j.cpc.2018.05.020} {\bibfield  {journal} {\bibinfo
  {journal} {Comput. Phys. Commun.}\ }\textbf {\bibinfo {volume} {232}},\
  \bibinfo {pages} {1} (\bibinfo {year} {2018})},\ \Eprint
  {https://arxiv.org/abs/1710.04903} {arXiv:1710.04903 [hep-ph]} \BibitemShut
  {NoStop}%
\bibitem [{\citenamefont {Huber}(2020)}]{Huber:2018ned}%
  \BibitemOpen
  \bibfield  {author} {\bibinfo {author} {\bibfnamefont {M.~Q.}\ \bibnamefont
  {Huber}},\ }\bibfield  {title} {\bibinfo {title} {{Nonperturbative properties
  of Yang\textendash{}Mills theories}},\ }\href
  {https://doi.org/10.1016/j.physrep.2020.04.004} {\bibfield  {journal}
  {\bibinfo  {journal} {Phys. Rept.}\ }\textbf {\bibinfo {volume} {879}},\
  \bibinfo {pages} {1} (\bibinfo {year} {2020})},\ \Eprint
  {https://arxiv.org/abs/1808.05227} {arXiv:1808.05227 [hep-ph]} \BibitemShut
  {NoStop}%
\bibitem [{\citenamefont {Fischer}(2019)}]{Fischer:2018sdj}%
  \BibitemOpen
  \bibfield  {author} {\bibinfo {author} {\bibfnamefont {C.~S.}\ \bibnamefont
  {Fischer}},\ }\bibfield  {title} {\bibinfo {title} {{QCD at finite
  temperature and chemical potential from Dyson\textendash{}Schwinger
  equations}},\ }\href {https://doi.org/10.1016/j.ppnp.2019.01.002} {\bibfield
  {journal} {\bibinfo  {journal} {Prog. Part. Nucl. Phys.}\ }\textbf {\bibinfo
  {volume} {105}},\ \bibinfo {pages} {1} (\bibinfo {year} {2019})},\ \Eprint
  {https://arxiv.org/abs/1810.12938} {arXiv:1810.12938 [hep-ph]} \BibitemShut
  {NoStop}%
\bibitem [{\citenamefont {Alkofer}\ \emph {et~al.}(2009)\citenamefont
  {Alkofer}, \citenamefont {Huber},\ and\ \citenamefont
  {Schwenzer}}]{Alkofer:2008nt}%
  \BibitemOpen
  \bibfield  {author} {\bibinfo {author} {\bibfnamefont {R.}~\bibnamefont
  {Alkofer}}, \bibinfo {author} {\bibfnamefont {M.~Q.}\ \bibnamefont {Huber}},\
  and\ \bibinfo {author} {\bibfnamefont {K.}~\bibnamefont {Schwenzer}},\
  }\bibfield  {title} {\bibinfo {title} {{Algorithmic derivation of
  Dyson-Schwinger Equations}},\ }\href
  {https://doi.org/10.1016/j.cpc.2008.12.009} {\bibfield  {journal} {\bibinfo
  {journal} {Comput. Phys. Commun.}\ }\textbf {\bibinfo {volume} {180}},\
  \bibinfo {pages} {965} (\bibinfo {year} {2009})},\ \Eprint
  {https://arxiv.org/abs/0808.2939} {arXiv:0808.2939 [hep-th]} \BibitemShut
  {NoStop}%
\bibitem [{\citenamefont {Huber}\ and\ \citenamefont
  {Braun}(2012)}]{Huber:2011qr}%
  \BibitemOpen
  \bibfield  {author} {\bibinfo {author} {\bibfnamefont {M.~Q.}\ \bibnamefont
  {Huber}}\ and\ \bibinfo {author} {\bibfnamefont {J.}~\bibnamefont {Braun}},\
  }\bibfield  {title} {\bibinfo {title} {{Algorithmic derivation of functional
  renormalization group equations and Dyson-Schwinger equations}},\ }\href
  {https://doi.org/10.1016/j.cpc.2012.01.014} {\bibfield  {journal} {\bibinfo
  {journal} {Comput.Phys.Commun.}\ }\textbf {\bibinfo {volume} {183}},\
  \bibinfo {pages} {1290} (\bibinfo {year} {2012})},\ \Eprint
  {https://arxiv.org/abs/1102.5307} {arXiv:1102.5307 [hep-th]} \BibitemShut
  {NoStop}%
\bibitem [{\citenamefont {Huber}\ \emph {et~al.}(2020)\citenamefont {Huber},
  \citenamefont {Cyrol},\ and\ \citenamefont {Pawlowski}}]{Huber_2020}%
  \BibitemOpen
  \bibfield  {author} {\bibinfo {author} {\bibfnamefont {M.~Q.}\ \bibnamefont
  {Huber}}, \bibinfo {author} {\bibfnamefont {A.~K.}\ \bibnamefont {Cyrol}},\
  and\ \bibinfo {author} {\bibfnamefont {J.~M.}\ \bibnamefont {Pawlowski}},\
  }\bibfield  {title} {\bibinfo {title} {Dofun 3.0: Functional equations in
  mathematica},\ }\href {https://doi.org/10.1016/j.cpc.2019.107058} {\bibfield
  {journal} {\bibinfo  {journal} {Computer Physics Communications}\ }\textbf
  {\bibinfo {volume} {248}},\ \bibinfo {pages} {107058} (\bibinfo {year}
  {2020})}\BibitemShut {NoStop}%
\bibitem [{\citenamefont {Huber}\ and\ \citenamefont
  {Cyrol}(2019)}]{github:DoFun}%
  \BibitemOpen
  \bibfield  {author} {\bibinfo {author} {\bibfnamefont {M.~Q.}\ \bibnamefont
  {Huber}}\ and\ \bibinfo {author} {\bibfnamefont {A.~K.}\ \bibnamefont
  {Cyrol}},\ }\href@noop {} {\bibinfo {title} {{DoFun GitHub Repository}}}
  (\bibinfo {year} {2019}),\ \bibinfo {note}
  {\url{https://github.com/markusqh/DoFun}}\BibitemShut {NoStop}%
\bibitem [{\citenamefont {Vermaseren}(2000)}]{Vermaseren:2000nd}%
  \BibitemOpen
  \bibfield  {author} {\bibinfo {author} {\bibfnamefont {J.~A.~M.}\
  \bibnamefont {Vermaseren}},\ }\bibfield  {title} {\bibinfo {title} {{New
  features of FORM}},\ }\href@noop {} {\  (\bibinfo {year} {2000})},\ \Eprint
  {https://arxiv.org/abs/math-ph/0010025} {arXiv:math-ph/0010025} \BibitemShut
  {NoStop}%
\bibitem [{\citenamefont {Vermaseren}(2016)}]{github:form}%
  \BibitemOpen
  \bibfield  {author} {\bibinfo {author} {\bibfnamefont {J.}~\bibnamefont
  {Vermaseren}},\ }\href@noop {} {\bibinfo {title} {{FORM GitHub Repository}}}
  (\bibinfo {year} {2016}),\ \bibinfo {note}
  {\url{https://github.com/vermaseren/form}}\BibitemShut {NoStop}%
\bibitem [{\citenamefont {Feng}\ and\ \citenamefont
  {Mertig}(2012{\natexlab{a}})}]{Feng:2012tk}%
  \BibitemOpen
  \bibfield  {author} {\bibinfo {author} {\bibfnamefont {F.}~\bibnamefont
  {Feng}}\ and\ \bibinfo {author} {\bibfnamefont {R.}~\bibnamefont {Mertig}},\
  }\bibfield  {title} {\bibinfo {title} {{FormLink/FeynCalcFormLink : Embedding
  FORM in Mathematica and FeynCalc}},\ }\href@noop {} {\  (\bibinfo {year}
  {2012}{\natexlab{a}})},\ \Eprint {https://arxiv.org/abs/1212.3522}
  {arXiv:1212.3522} \BibitemShut {NoStop}%
\bibitem [{\citenamefont {Feng}\ and\ \citenamefont
  {Mertig}(2012{\natexlab{b}})}]{github:FormLink}%
  \BibitemOpen
  \bibfield  {author} {\bibinfo {author} {\bibfnamefont {F.}~\bibnamefont
  {Feng}}\ and\ \bibinfo {author} {\bibfnamefont {R.}~\bibnamefont {Mertig}},\
  }\href@noop {} {\bibinfo {title} {{FormLink/FeynCalcFormLink GitHub
  Repository}}} (\bibinfo {year} {2012}{\natexlab{b}}),\ \bibinfo {note}
  {\url{https://github.com/FormLink/formlink}}\BibitemShut {NoStop}%
\bibitem [{\citenamefont {Cyrol}\ \emph {et~al.}(2017)\citenamefont {Cyrol},
  \citenamefont {Mitter},\ and\ \citenamefont {Strodthoff}}]{Cyrol:2016zqb}%
  \BibitemOpen
  \bibfield  {author} {\bibinfo {author} {\bibfnamefont {A.~K.}\ \bibnamefont
  {Cyrol}}, \bibinfo {author} {\bibfnamefont {M.}~\bibnamefont {Mitter}},\ and\
  \bibinfo {author} {\bibfnamefont {N.}~\bibnamefont {Strodthoff}},\ }\bibfield
   {title} {\bibinfo {title} {{FormTracer - A Mathematica Tracing Package Using
  FORM}},\ }\href {https://doi.org/10.1016/j.cpc.2017.05.024} {\bibfield
  {journal} {\bibinfo  {journal} {Comput. Phys. Commun.}\ }\textbf {\bibinfo
  {volume} {C219}},\ \bibinfo {pages} {346} (\bibinfo {year} {2017})},\ \Eprint
  {https://arxiv.org/abs/1610.09331} {arXiv:1610.09331 [hep-ph]} \BibitemShut
  {NoStop}%
\bibitem [{\citenamefont {Cyrol}\ \emph {et~al.}(2016)\citenamefont {Cyrol},
  \citenamefont {Mitter}, \citenamefont {Pawlowski},\ and\ \citenamefont
  {Strodthoff}}]{github:FormTracer}%
  \BibitemOpen
  \bibfield  {author} {\bibinfo {author} {\bibfnamefont {A.~K.}\ \bibnamefont
  {Cyrol}}, \bibinfo {author} {\bibfnamefont {M.}~\bibnamefont {Mitter}},
  \bibinfo {author} {\bibfnamefont {J.~M.}\ \bibnamefont {Pawlowski}},\ and\
  \bibinfo {author} {\bibfnamefont {N.}~\bibnamefont {Strodthoff}},\
  }\href@noop {} {\bibinfo {title} {{FormTracer GitHub Repository}}} (\bibinfo
  {year} {2016}),\ \bibinfo {note}
  {\url{https://github.com/FormTracer/FormTracer}}\BibitemShut {NoStop}%
\bibitem [{\citenamefont {Litim}\ and\ \citenamefont
  {Steudtner}(2020{\natexlab{a}})}]{Litim:2020jvl}%
  \BibitemOpen
  \bibfield  {author} {\bibinfo {author} {\bibfnamefont {D.~F.}\ \bibnamefont
  {Litim}}\ and\ \bibinfo {author} {\bibfnamefont {T.}~\bibnamefont
  {Steudtner}},\ }\bibfield  {title} {\bibinfo {title} {{ARGES -- Advanced
  Renormalisation Group Equation Simplifier}},\ }\href@noop {} {\  (\bibinfo
  {year} {2020}{\natexlab{a}})},\ \Eprint {https://arxiv.org/abs/2012.12955}
  {arXiv:2012.12955 [hep-ph]} \BibitemShut {NoStop}%
\bibitem [{\citenamefont {Litim}\ and\ \citenamefont
  {Steudtner}(2020{\natexlab{b}})}]{github:ARGES}%
  \BibitemOpen
  \bibfield  {author} {\bibinfo {author} {\bibfnamefont {D.~F.}\ \bibnamefont
  {Litim}}\ and\ \bibinfo {author} {\bibfnamefont {T.}~\bibnamefont
  {Steudtner}},\ }\href@noop {} {\bibinfo {title} {{ARGES GitHub Repository}}}
  (\bibinfo {year} {2020}{\natexlab{b}}),\ \bibinfo {note}
  {\url{https://github.com/TomSteu/ARGES}}\BibitemShut {NoStop}%
\bibitem [{\citenamefont {Dyson}(1949)}]{Dyson:1949ha}%
  \BibitemOpen
  \bibfield  {author} {\bibinfo {author} {\bibfnamefont {F.}~\bibnamefont
  {Dyson}},\ }\bibfield  {title} {\bibinfo {title} {{The S matrix in quantum
  electrodynamics}},\ }\href {https://doi.org/10.1103/PhysRev.75.1736}
  {\bibfield  {journal} {\bibinfo  {journal} {Phys. Rev.}\ }\textbf {\bibinfo
  {volume} {75}},\ \bibinfo {pages} {1736} (\bibinfo {year}
  {1949})}\BibitemShut {NoStop}%
\bibitem [{\citenamefont {Schwinger}(1951)}]{Schwinger:1951ex}%
  \BibitemOpen
  \bibfield  {author} {\bibinfo {author} {\bibfnamefont {J.~S.}\ \bibnamefont
  {Schwinger}},\ }\bibfield  {title} {\bibinfo {title} {{On the Green's
  functions of quantized fields. 1.}},\ }\href
  {https://doi.org/10.1073/pnas.37.7.452} {\bibfield  {journal} {\bibinfo
  {journal} {Proc. Nat. Acad. Sci.}\ }\textbf {\bibinfo {volume} {37}},\
  \bibinfo {pages} {452} (\bibinfo {year} {1951})}\BibitemShut {NoStop}%
\bibitem [{\citenamefont {Symanzik}(1970)}]{Symanzik:1970rt}%
  \BibitemOpen
  \bibfield  {author} {\bibinfo {author} {\bibfnamefont {K.}~\bibnamefont
  {Symanzik}},\ }\bibfield  {title} {\bibinfo {title} {{Small distance behavior
  in field theory and power counting}},\ }\href
  {https://doi.org/10.1007/BF01649434} {\bibfield  {journal} {\bibinfo
  {journal} {Commun. Math. Phys.}\ }\textbf {\bibinfo {volume} {18}},\ \bibinfo
  {pages} {227} (\bibinfo {year} {1970})}\BibitemShut {NoStop}%
\bibitem [{\citenamefont {Wetterich}(1993)}]{Wetterich:1992yh}%
  \BibitemOpen
  \bibfield  {author} {\bibinfo {author} {\bibfnamefont {C.}~\bibnamefont
  {Wetterich}},\ }\bibfield  {title} {\bibinfo {title} {{Exact evolution
  equation for the effective potential}},\ }\href
  {https://doi.org/10.1016/0370-2693(93)90726-X} {\bibfield  {journal}
  {\bibinfo  {journal} {Phys.Lett.}\ }\textbf {\bibinfo {volume} {B301}},\
  \bibinfo {pages} {90} (\bibinfo {year} {1993})}\BibitemShut {NoStop}%
\bibitem [{\citenamefont {Ellwanger}(1993)}]{Ellwanger:1993mw}%
  \BibitemOpen
  \bibfield  {author} {\bibinfo {author} {\bibfnamefont {U.}~\bibnamefont
  {Ellwanger}},\ }\bibfield  {title} {\bibinfo {title} {{FLow equations for N
  point functions and bound states}},\ }\href
  {https://doi.org/10.1007/BF01555911} {\ ,\ \bibinfo {pages} {206} (\bibinfo
  {year} {1993})},\ \Eprint {https://arxiv.org/abs/hep-ph/9308260}
  {arXiv:hep-ph/9308260} \BibitemShut {NoStop}%
\bibitem [{\citenamefont {Bonini}\ \emph {et~al.}(1993)\citenamefont {Bonini},
  \citenamefont {D'Attanasio},\ and\ \citenamefont
  {Marchesini}}]{Bonini:1992vh}%
  \BibitemOpen
  \bibfield  {author} {\bibinfo {author} {\bibfnamefont {M.}~\bibnamefont
  {Bonini}}, \bibinfo {author} {\bibfnamefont {M.}~\bibnamefont
  {D'Attanasio}},\ and\ \bibinfo {author} {\bibfnamefont {G.}~\bibnamefont
  {Marchesini}},\ }\bibfield  {title} {\bibinfo {title} {{Perturbative
  renormalization and infrared finiteness in the Wilson renormalization group:
  The Massless scalar case}},\ }\href
  {https://doi.org/10.1016/0550-3213(93)90588-G} {\bibfield  {journal}
  {\bibinfo  {journal} {Nucl. Phys. B}\ }\textbf {\bibinfo {volume} {409}},\
  \bibinfo {pages} {441} (\bibinfo {year} {1993})},\ \Eprint
  {https://arxiv.org/abs/hep-th/9301114} {arXiv:hep-th/9301114} \BibitemShut
  {NoStop}%
\bibitem [{\citenamefont {Morris}(1994)}]{Morris:1993qb}%
  \BibitemOpen
  \bibfield  {author} {\bibinfo {author} {\bibfnamefont {T.~R.}\ \bibnamefont
  {Morris}},\ }\bibfield  {title} {\bibinfo {title} {{The Exact renormalization
  group and approximate solutions}},\ }\href
  {https://doi.org/10.1142/S0217751X94000972} {\bibfield  {journal} {\bibinfo
  {journal} {Int. J. Mod. Phys.}\ }\textbf {\bibinfo {volume} {A9}},\ \bibinfo
  {pages} {2411} (\bibinfo {year} {1994})},\ \Eprint
  {https://arxiv.org/abs/hep-ph/9308265} {arXiv:hep-ph/9308265} \BibitemShut
  {NoStop}%
\bibitem [{\citenamefont {Becchi}\ \emph {et~al.}(1976)\citenamefont {Becchi},
  \citenamefont {Rouet},\ and\ \citenamefont {Stora}}]{Becchi:1975nq}%
  \BibitemOpen
  \bibfield  {author} {\bibinfo {author} {\bibfnamefont {C.}~\bibnamefont
  {Becchi}}, \bibinfo {author} {\bibfnamefont {A.}~\bibnamefont {Rouet}},\ and\
  \bibinfo {author} {\bibfnamefont {R.}~\bibnamefont {Stora}},\ }\bibfield
  {title} {\bibinfo {title} {{Renormalization of Gauge Theories}},\ }\href
  {https://doi.org/10.1016/0003-4916(76)90156-1} {\bibfield  {journal}
  {\bibinfo  {journal} {Annals Phys.}\ }\textbf {\bibinfo {volume} {98}},\
  \bibinfo {pages} {287} (\bibinfo {year} {1976})}\BibitemShut {NoStop}%
\bibitem [{\citenamefont {Tyutin}(1975)}]{Tyutin:1975qk}%
  \BibitemOpen
  \bibfield  {author} {\bibinfo {author} {\bibfnamefont {I.}~\bibnamefont
  {Tyutin}},\ }\bibfield  {title} {\bibinfo {title} {{Gauge Invariance in Field
  Theory and Statistical Physics in Operator Formalism}},\ }\href@noop {} {\
  (\bibinfo {year} {1975})},\ \Eprint {https://arxiv.org/abs/0812.0580}
  {arXiv:0812.0580 [hep-th]} \BibitemShut {NoStop}%
\bibitem [{\citenamefont {Zinn-Justin}(1975)}]{ZinnJustin:1974mc}%
  \BibitemOpen
  \bibfield  {author} {\bibinfo {author} {\bibfnamefont {J.}~\bibnamefont
  {Zinn-Justin}},\ }\bibfield  {title} {\bibinfo {title} {{Renormalization of
  Gauge Theories}},\ }\href {https://doi.org/10.1007/3-540-07160-1_1}
  {\bibfield  {journal} {\bibinfo  {journal} {Lect. Notes Phys.}\ }\textbf
  {\bibinfo {volume} {37}},\ \bibinfo {pages} {1} (\bibinfo {year}
  {1975})}\BibitemShut {NoStop}%
\bibitem [{\citenamefont {Zinn-Justin}(1999)}]{ZinnJustin:1999ze}%
  \BibitemOpen
  \bibfield  {author} {\bibinfo {author} {\bibfnamefont {J.}~\bibnamefont
  {Zinn-Justin}},\ }\bibfield  {title} {\bibinfo {title} {{Renormalization of
  gauge theories and master equation}},\ }\href
  {https://doi.org/10.1142/S0217732399001322} {\bibfield  {journal} {\bibinfo
  {journal} {Mod. Phys. Lett. A}\ }\textbf {\bibinfo {volume} {14}},\ \bibinfo
  {pages} {1227} (\bibinfo {year} {1999})},\ \Eprint
  {https://arxiv.org/abs/hep-th/9906115} {arXiv:hep-th/9906115} \BibitemShut
  {NoStop}%
\end{thebibliography}%

\end{document}